\documentclass[cernpreprint, atlasdraft=false, texlive=2023, UKenglish, orcidlogo, texmf, orcidlogo]{atlasdoc}
 
\usepackage{atlaspackage}
\usepackage{atlasbiblatex}
 
\usepackage{atlasphysics}
\usepackage{cancel}
 
\graphicspath{{logos/}{figures/}}
 
\usepackage{ANA-HIGG-2020-18-PAPER-defs}
\usepackage{lscape}
 
\usepackage{pdflscape}
 
\usepackage{multirow}

\AtlasTitle{Model-independent search for the presence of new physics in
events including $H \rightarrow \gamma\gamma$ with $\sqrt{s} = 13$ TeV $pp$ data recorded by the ATLAS detector at the LHC}

\AtlasAbstract{
A model-independent search for new physics leading to final states containing a Higgs boson, with a mass of 125.09 GeV, decaying to a pair of photons is performed with 139 fb$^{-1}$ of $\sqrt{s} = 13$ TeV $pp$ collision data recorded by the ATLAS detector at the Large Hadron Collider at CERN. This search examines 22 final states categorized by the objects that are produced in association with the Higgs   boson. These objects include isolated electrons or muons, hadronically decaying $\tau$-leptons, additional photons, missing transverse momentum, and hadronic jets, as well as jets that are tagged   as containing a $b$-hadron. No significant excesses above Standard Model expectations are observed and limits on the production cross section at 95\% confidence level are set. Detector efficiencies are reported for all 22 signal regions, which can be used to convert detector-level cross-section limits reported in this paper to particle-level cross-section constraints.
}

\AtlasRefCode{HIGG-2020-18}
 
\PreprintIdNumber{CERN-EP-2022-232}

\AtlasJournalRef{JHEP 07 (2023) 176}
\AtlasDOI{10.1007/JHEP07(2023)176}

\hypersetup{pdftitle={ATLAS document},pdfauthor={The ATLAS Collaboration}}
 
\begin{document}
 
\maketitle
 
\tableofcontents
 
\section{Introduction}

% The next lines are included from the .//sections/introduction.tex input file
The discovery of the Higgs boson by the ATLAS and CMS experiments at the CERN Large Hadron Collider (LHC)~\cite{HIGG-2012-27,CMS-HIG-12-028} opened up new avenues to search for physics beyond the Standard Model (BSM physics). The properties of the Higgs boson, including its inclusive production cross sections and decay rates, as well as its differential cross sections, may be altered by the presence of new interactions and new particles beyond the SM. In addition, the Higgs boson may also be produced in BSM processes as a decay product of unknown new particles, which could give rise to deviations of Higgs boson production rates from SM predictions in exclusive final states. For example, the Higgs boson can be produced in the decays of supersymmetric particles predicted by weak-scale supersymmetry (SUSY) models~\cite{Golfand:1971iw,Volkov:1973ix,Wess:1974tw,Wess:1974jb,Ferrara:1974pu,Salam:1974ig}, in exotic decays of top quarks in models such as those having flavour-changing neutral currents (FCNCs)~\cite{Aguilar-Saavedra:2002phh,Bejar:2000ub,Guasch:1999jp,Cao:2007dk,Eilam:2001dh}, or in the decay of the partner of the top quark or bottom quark
as predicted by models of vector-like quarks~\cite{delAguila:1982fs,Aguilar-Saavedra:2009xmz}. So far, observations of Higgs boson production and decay properties at the LHC are consistent with SM predictions~\cite{ATLAS:2022vkf,CMS:2022dwd}, and direct searches targeting the aforementioned BSM models also reported no significant deviations from the SM expectation. Most of these analyses are designed with certain model assumptions, either targeting a specific SM Higgs boson production mode or optimized for a given BSM process involving the production of the Higgs boson.
 
This paper reports the results of a model-independent search for excesses in the production rate of the Higgs boson with a mass of 125.09~\GeV\ in many distinct signal regions with the ATLAS detector at the LHC. The regions are defined by the presence and kinematic properties of objects produced in association with the Higgs boson. While a large fraction of such signals may populate signal regions probed by SM Higgs boson measurements, such as measurements of \enquote{simplified template cross sections}~\cite{deFlorian:2016spz,HIGG-2020-16} and fiducial and differential cross-section measurements~\cite{HIGG-2022-04}, unknown BSM production of the Higgs boson could result in events that do not enter signal regions optimized for existing SM measurements. The signal regions in this search are chosen to cover a wide range of signatures defined by particles accompanying the production of the Higgs boson, and they are defined with simple and mostly inclusive selections in order to minimize model dependence.
 
In this search the Higgs boson is reconstructed via its diphoton decay channel, which has a predicted branching ratio of ($0.227 \pm 0.007$)\%~\cite{deFlorian:2016spz} in the SM. While the rate of \Hgg\ decay is much lower than that of several other Higgs boson decay channels, 
the sensitivity of an analysis targeting this channel can be competitive with that of analyses targeting higher-rate channels.
This is primarily because the diphoton background to the Higgs boson signal is relatively small and because of
better experimental resolution for the reconstructed resonance mass, especially in comparison with hadronic decay modes of the Higgs boson.
Another advantage of performing this model-independent search in the \Hgg\ channel is that the same background modelling strategy can be applied to most signal regions; the signal and background can be determined by fitting analytic functions to the diphoton invariant mass (\mgg) distribution in data. This fit-based background estimation method breaks down only when the signal region does not have a sufficient number of events to constrain the analytic background function. 
 
If data in the signal regions are found to be consistent with the SM expectations, including contributions from resonant Higgs boson production, limits are set on the BSM production cross section of the Higgs boson in detector-level signal regions. To enable the use of these detector-level cross-section limits in constraining BSM physics models, the detector efficiencies for different signal regions are evaluated. These efficiencies depend on the kinematics of the final-state particles and therefore vary between different BSM models. As a result, this paper reports a range of efficiency values for each signal region, evaluated from Monte Carlo simulation of several benchmark BSM models. Correlations in the observed limits between signal regions with a large number of events are also reported.

This paper is organized as follows. Section~\ref{sec:detector} gives a brief description of the ATLAS detector.  Section~\ref{sec:monte} presents the simulated SM samples used to describe the backgrounds, and also the new-physics models used to determine the expected sensitivity of the search in each signal region.  Section~\ref{sec:data} describes the reconstructed data objects and subsequent selections used in the analysis. Section \ref{sec:region} discusses in detail the signal regions defined for the search. Section~\ref{sec:modeling} describes the modelling of signal and background in each search region. Section~\ref{sec:systematics} addresses systematic uncertainties related to the search.  Section~\ref{sec:result} presents the results of this search. Finally, Section~\ref{sec:conclusion} summarizes the conclusions of the paper.

% End of text imported from the .//sections/introduction.tex input file

\section{ATLAS detector}
\label{sec:detector}

% The next lines are included from the .//sections/atlas-detector.tex input file
The ATLAS detector~\cite{PERF-2007-01} at the LHC covers nearly the entire solid angle around the collision point.\footnote{\AtlasCoordFootnote}
It consists of an inner tracking detector surrounded by a thin superconducting solenoid, electromagnetic and hadronic calorimeters,
and a muon spectrometer incorporating three large superconducting toroidal magnet systems.
The inner-detector system is immersed in a \SI{2}{\tesla} axial magnetic field
and provides charged-particle tracking in the range \(|\eta| < 2.5\).
 
The fine-granularity silicon pixel detector covers the vertex region and typically provides four measurements per track,
the first hit normally being in the insertable B-layer installed before Run~2~\cite{ATLASDET-IBL,PIX-2018-001}.
The pixel detector is followed by the silicon microstrip tracker, which usually provides eight measurements per track.
These silicon detectors are complemented by the transition radiation tracker (TRT),
which enables radially extended track reconstruction up to \(|\eta| = 2.0\).
The TRT also provides electron identification information
based on the fraction of hits (typically 30 hits in total) above a higher energy-deposit threshold corresponding to transition radiation.
 
The calorimeter system covers the pseudorapidity range \(|\eta| < 4.9\). Within the region \(|\eta|< 3.2\), electromagnetic (EM) calorimetry is provided by barrel and
endcap lead/liquid-argon (LAr) calorimeters,
with an additional thin LAr presampler covering \(|\eta| < 1.8\)
to correct for energy loss in material upstream of the calorimeters.
Hadronic calorimetry is provided by a steel/scintillator-tile calorimeter,
segmented into three barrel structures within \(|\eta| < 1.7\), and two copper/LAr hadronic endcap calorimeters.
The solid angle coverage is completed with forward copper/LAr and tungsten/LAr calorimeter modules
optimized for electromagnetic and hadronic measurements, respectively.
 
The muon spectrometer comprises separate trigger and
high-precision tracking chambers measuring the deflection of muons in a magnetic field generated by the superconducting air-core toroids.
The field integral of the toroids ranges between \num{2.0} and \SI{6.0}{\tesla\metre}
across most of the detector.
A set of precision chambers covers the region \(|\eta| < 2.7\) with three layers of monitored drift tubes,
complemented by cathode-strip chambers in the forward region, where the background is highest.
The muon trigger system covers the range \(|\eta| < 2.4\) with resistive-plate chambers in the barrel, and thin-gap chambers in the endcap regions.
 
Interesting events are selected to be recorded by the first-level trigger system implemented in custom hardware,
followed by selections made by algorithms implemented in software in the high-level trigger~\cite{TRIG-2016-01}.
The first-level trigger accepts events from the \SI{40}{\MHz} bunch crossings at a rate below \SI{100}{\kHz},
which the high-level trigger reduces in order to record events to disk at about \SI{1}{\kHz}.
An extensive software suite~\cite{ATL-SOFT-PUB-2021-001} is used in data simulation, in the reconstruction and analysis of real and simulated data, in detector operations, and in the trigger and data acquisition systems of the experiment.

% End of text imported from the .//sections/atlas-detector.tex input file

\section{Data and Monte Carlo samples}
\label{sec:monte}

% The next lines are included from the .//sections/monte.tex input file

\subsection{Data sample}
This study uses the full Run-2 data set of $\sqrt{s}=13$~\TeV\ proton--proton collisions recorded by the ATLAS detector during 2015--2018. After data quality requirements~\cite{DAPR-2018-01} are applied to ensure the good working condition of all detector components, the data set amounts to an integrated luminosity of $139.0 \pm 2.4$~fb$^{-1}$~\cite{ATLAS-CONF-2019-021,LUCID2}. The mean number of interactions per bunch crossing was $\langle \mu \rangle=33.7$  on average during Run~2.
 
The data sample was selected using a combination of diphoton and single-photon triggers. The transverse energy thresholds of the diphoton trigger were 35~\GeV\ and 25~\GeV\ for the leading and subleading photon candidates, respectively~\cite{TRIG-2018-05}. The diphoton trigger applies photon identification selections based on calorimeter shower shape variables. In 2015--2016, a loose photon identification requirement was used, and in 2017--2018, this requirement was tightened to cope with higher instantaneous luminosity. The single-photon trigger had a transverse energy threshold of 120~\GeV\ for the leading photon in data collected between 2015 and 2017, and the threshold was increased to 140~\GeV\ for data collected in 2018. The photon candidate used in the trigger decision is required to satisfy the loose photon identification criteria. The use of the single-photon trigger is found to improve the trigger efficiency for high-\pt\ Higgs boson candidates. Typically, the trigger efficiency is greater than 98\% for events that pass the diphoton event selection described below.

\subsection{Monte Carlo samples}
 
Monte Carlo (MC) simulation samples are used to determine the SM Higgs boson event yield, test the continuum background modelling, and estimate detector efficiencies for various signal regions. These simulated samples include three categories of processes: SM Higgs boson production, SM diphoton continuum production, and BSM production of the Higgs boson with a mass of 125.09 GeV.
 
\begin{table}[!htbp]
\centering
\caption{Configuration of MC simulation of SM Higgs boson production and the predicted cross sections at $\sqrt{s} = 13$~\TeV. The generator set-up includes the software programs used for the matrix element (ME) calculation and parton showering (PS). The parton distribution function (PDF) is given separately for the ME calculation and the parton showering, along with the set of tuned parameters (Tune) used in modelling the underlying event (UE). }
\scriptsize
\begin{tabular}{lcccc}
 
\toprule
Process & Generator & PDF  & PDF, Tune & Cross section [pb] \\
&  (ME+PS)  &   (ME)       &  (PS, UE)      &($\sqrt{s} = 13$~\TeV)\\

\midrule
ggF			& \POWHEG NNLOPS + \Pythia[8] 	& PDF4LHC15 & CTEQ6, AZNLO & 48.5\phantom{00}\\
VBF              		& \POWHEGBOX + \Pythia[8]          	& PDF4LHC15 & CTEQ6, AZNLO & 3.78\\
$WH$           		& \POWHEGBOX + \Pythia[8]           & PDF4LHC15 & CTEQ6, AZNLO & 1.37\\
$q\bar{q}\to ZH$ 	& \POWHEGBOX + \Pythia[8]           & PDF4LHC15 & CTEQ6, AZNLO & 0.76\\
$gg\to ZH$       	& \POWHEGBOX + \Pythia[8]           & PDF4LHC15 & CTEQ6, AZNLO & 0.12\\
$t\bar{t}H$      	& \POWHEGBOX + \Pythia[8]           & PDF4LHC15 & NNPDF2.3, A14 & 0.51\\
$b\bar{b}H$      	& \POWHEGBOX + \Pythia[8]          	& PDF4LHC15 & NNPDF2.3, A14 & 0.49\\
$tH bj$ 			& aMC@NLO + \Pythia[8] 		& NNPDF3.0 	& NNPDF2.3, A14 & \phantom{0}0.074\\
$tWH$ 			& aMC@NLO + \Pythia[8]   		& NNPDF3.0 	& NNPDF2.3, A14 & \phantom{0}0.015 \\
 
\bottomrule
\end{tabular}
\label{tab:samples:nominal_mc}
\end{table}
 
MC samples generated for SM Higgs boson production are summarized in Table~\ref{tab:samples:nominal_mc}. The gluon--gluon fusion (\ggF),
vector-boson fusion (VBF), and \VH\ production modes were generated using \POWHEG~NNLOPS or \POWHEGBOX~\cite{Hamilton:2013fea,Nason:2004rx,Frixione:2007vw,Alioli:2010xd,Nason:2009ai,Mimasu:2015nqa,Campbell:2012am,Luisoni:2013kna,Hartanto:2015uka},
with the \PDFforLHC[15] PDF set~\cite{Butterworth:2015oua}, and interfaced to \PYTHIA[8]~\cite{pythia8,Sjostrand:2014zea} for simulation of parton showering,
hadronization and the underlying event, using a set of data-tuned parameters called the AZNLO tune~\cite{Aad:2014xaa}. The \ggF\ simulation achieves
NNLO accuracy for arbitrary inclusive $gg\to H$ observables by
reweighting the Higgs boson rapidity spectrum in
{\textsc Hj-MiNLO}~\cite{Hamilton:2012np,Campbell:2012am,Hamilton:2012rf}
to that of HNNLO~\cite{Catani:2007vq}. The transverse momentum spectrum of the Higgs boson obtained
with this sample is found to be compatible with the fixed-order HNNLO calculation and the \HRES[2.3]
calculation~\cite{Bozzi:2005wk,deFlorian:2011xf} performing resummation at
next-to-next-to-leading-logarithm accuracy matched to a NNLO fixed-order calculation (NNLL+NNLO).
The VBF production mode was simulated at next-to-leading-order (NLO) accuracy in QCD. The simulation of \WH\ and \qqtoZH\ production is accurate to NLO in QCD with up to one extra jet in the event, while the simulation of the \ggtoZH\ process was performed at leading order in QCD.
The \ttH\ and \bbH\ processes were modelled using the \POWHEGBOX[v2]~\cite{Frixione:2007nw,Nason:2004rx,Frixione:2007vw,Alioli:2010xd,Hartanto:2015uka}
generator, which provides matrix elements at NLO in the strong coupling constant \alphas in the
five-flavour scheme with the \NNPDF[2.3lo]~\cite{Ball:2014uwa} PDF set. For the \ttH\ sample, the functional form of the
renormalization and factorization scale was set to $\sqrt[3]{m_\text{T}(t)\cdot m_\text{T}(\bar{t}) \cdot m_\text{T}(H)}$\footnote{$m_\text{T}$ denotes the transverse mass
of a particle, defined as $m_\text{T}=\sqrt{m^2 + p_\text{T}^2}$ where $m$ and $p_\text{T}$
are its mass and transverse momentum, respectively.}~\cite{deFlorian:2016spz}. The generator was interfaced to \PYTHIA[8.230]~\cite{Sjostrand:2014zea},
which used the A14 tune~\cite{ATL-PHYS-PUB-2014-021} and
the \NNPDF[2.3lo]~\cite{Ball:2014uwa} PDF set. The decays of bottom and charm hadrons were simulated
using the \EVTGEN[1.6.0] program~\cite{EvtGen}.
All the SM Higgs boson samples are normalized to cross-section values recommended by Ref.~\cite{deFlorian:2016spz}, which are shown in Table~\ref{tab:samples:nominal_mc}.

To study the modelling of continuum background, diphoton production event samples were simulated. Besides inclusive diphoton production, specific diphoton processes were also generated for signal regions where electroweak or top-associated production is enhanced. Inclusive diphoton production and diphoton production in association with a vector boson ($V\gamma\gamma$) were simulated with the \Sherpa[2.2.4]~\cite{Bothmann:2019yzt} generator. Matrix elements (MEs) for up to three additional partons were calculated at LO accuracy. For the inclusive diphoton sample, NLO accuracy is reached for MEs including up to one additional parton.  The Comix~\cite{Gleisberg:2008fv} and
\OPENLOOPS~\cite{Buccioni:2019sur,Cascioli:2011va,Denner:2016kdg} libraries were used in these calculations, which were matched with the \Sherpa\ parton shower~\cite{Schumann:2007mg} using the MEPS@NLO
prescription~\cite{Hoeche:2011fd,Hoeche:2012yf,Catani:2001cc,Hoeche:2009rj}
with a dynamic merging cut~\cite{Siegert:2016bre} of 10~\GeV.  A smooth-cone isolation requirement~\cite{Frixione:1998jh} was applied in the event generation. Samples were generated using the
\NNPDF[3.0nnlo] PDF set~\cite{Ball:2014uwa},
along with the dedicated set of tuned parton-shower parameters developed by the \Sherpa\ authors.
Diphoton production in association with a top-quark pair ($t\bar{t}\gamma\gamma$) was modelled by the \MGNLO[2.3.3]
generator at LO with the \NNPDF[2.3lo]~\cite{Ball:2012cx} PDF set, and the events were interfaced with \PYTHIA[8.212] for parton-shower and underlying-event simulation.
 
Several \BSM\ processes were simulated to demonstrate the coverage of the signal regions and estimate the model dependence of the efficiency of the fiducial selections. These samples are summarized in Table~\ref{tab:samples:bsm} and are described as follows.
 
\begin{enumerate}
\item BSM Higgs boson production characterized by the presence of additional charged or neutral leptons:
\begin{itemize}
\item   \ConeNtwo: the production of a neutralino--chargino pair that decays to final states involving $W$, $Z$ and Higgs bosons.
\item \ConeCone: the production of a chargino pair that decays to final states involving Higgs bosons and leptons through an $R$-parity-violating interaction.
\end{itemize}
\item BSM Higgs boson production characterized by the presence of additional heavy-flavour jets are represented by
\begin{itemize}
\item $\tilde{t}\tilde{t}$, $\tilde{b}\tilde{b}$: stop-pair production  and sbottom-pair production  processes. In the subsequent decays of the stop or sbottom particles, the Higgs boson could be produced in the decays of next-to-lightest supersymmetric particles.
\end{itemize}
\item BSM Higgs boson production characterized by the presence of an additional top quark is represented by
\begin{itemize}
\item \fcnc: pair-produced top quarks that decay into final states involving Higgs bosons, through a FCNC interaction.
\end{itemize}
\item BSM Higgs boson production characterized by the presence of an additional photon is represented by
\begin{itemize}
\item \exotic: a massive neutral boson ($\chi$) decaying into a photon and a Higgs boson, a process predicted by extensions of the Higgs sector or gauge sectors~\cite{ArkaniHamed:2002qy,Agashe:2004rs, Randall:1999vf}.
\end{itemize}
\end{enumerate}
 
These \BSM\ processes are studied in various dedicated ATLAS searches that may use Higgs boson decay channels other than \Hgg. The specific process, model parameters, expected cross section, and a reference to the dedicated search, are summarized in Table~\ref{tab:samples:bsm}. These BSM MC samples were generated with the same set-up as used in the dedicated searches, except that the Higgs boson is configured to decay into a pair of photons.

The SUSY events were generated with up to two additional partons in the matrix element using \MGNLO[2.6.2] at LO in QCD with the \NNPDF[3.0lo] PDF set and CKKW-L merging scheme. Parton showering and hadronization were handled by \PYTHIA[8.230] with the A14 tune, using the \NNPDF[2.3lo] PDF set.
 
MC samples were generated for the BSM process where a massive neutral resonance is produced and decays into a Higgs boson and a photon (\exotic). This process could result in a final state with at least three photons. The event generation used the same set-up as for the SUSY samples, except that \PYTHIA[8.235] was used to model the parton shower and underlying event.
 
A \ttbar\ sample with FCNC-induced top-quark decays is also considered. In this process, one top quark decays into a Higgs boson and an up-type quark through a FCNC interaction, while the other top quark decays into a $b$-quark and a $W$ boson. These FCNC samples were simulated using \MGNLO[2.4.3]~\cite{Alwall:2014hca} interfaced to \PYTHIA[8.212]~\cite{pythia8} with the A14 tune for the modelling of parton showers, hadronization and the underlying event. The FCNC samples were generated separately for the up-quark and charm-quark final states.

\begin{table}[!htbp]
\centering
\caption{Summary of processes, model parameters, and cross sections for the BSM samples used in this paper. A reference to their dedicated ATLAS search is also included. The cross sections quoted include the SM branching ratio for $H\to \gamma\gamma$. The \exotic\   process is expected in multiple BSM models, so no particular model is used for the cross-section calculation.}
\scriptsize
\begin{tabular}{lccc}
\toprule
Process with $H \rightarrow \gamma\gamma$ &  Signal mass & Cross section [pb] & Ref. \\
\midrule
\ConeNtwo  &  $m(\tilde{\chi}_1^\pm/\tilde{\chi}_2^0), m(\tilde{\chi}_1^0) = 150, 0.5$~\GeV  & 5.18\phantom{000}  & ~\cite{SUSY-2018-23} \\ 
\ConeNtwo  &  $m(\tilde{\chi}_1^\pm/\tilde{\chi}_2^0), m(\tilde{\chi}_1^0) = 300, 0.5$~\GeV  & 0.39\phantom{000} & ~\cite{SUSY-2018-23}\\ 
\ConeCone &  $m(\tilde{\chi}_1^\pm) = 150$~\GeV  & 2.61\phantom{000} & ~\cite{SUSY-2018-36}\\ 
\ConeCone   &  $m(\tilde{\chi}_1^\pm) = 300$~\GeV  & 0.19\phantom{000} & ~\cite{SUSY-2018-36} \\ 
\ConeNone  &  $m(\tilde{\chi}_1^\pm/\tilde{\chi}_1^0) = 200$~\GeV  & 1.81\phantom{000} &~\cite{SUSY-2018-36}\\ 
\ConeNone  & $m(\tilde{\chi}_1^\pm/\tilde{\chi}_1^0) = 400$~\GeV  & 0.12\phantom{000} &~\cite{SUSY-2018-36}\\ 
\stopTwo&  $m(\tilde{t}_2), m(\tilde{t}_1), m(\tilde{\chi}_2^0) = 500, 340, 300$~\GeV & 0.61\phantom{000} & ~\cite{SUSY-2018-21} \\ 
\sbottom  &  $m(\tilde{b}), m(\tilde{\chi}_2^0), m(\tilde{\chi}_1^0) = 500, 180, 50$~\GeV & 0.61\phantom{000} & ~\cite{SUSY-2018-31}\\ 
\sbottom  &  $ m(\tilde{b}), m(\tilde{\chi}_2^0), m(\tilde{\chi}_1^0) = 1000, 205, 60$~\GeV & 0.0068\phantom{0} & ~\cite{SUSY-2018-31}\\ 
\sbottom &  $m(\tilde{b}), m(\tilde{\chi}_2^0), m(\tilde{\chi}_1^0) = 1200, 205, 60$~\GeV & 0.0017\phantom{0} & ~\cite{SUSY-2018-31} \\
\exotic &  $m(\chi) = 200$~\GeV& - &~\cite{HDBS-2018-17}\\
\exotic &  $m(\chi) = 500$~\GeV& - & ~\cite{HDBS-2018-17}\\
\fcnc  &-& 0.00123 & ~\cite{HIGG-2016-26}\\ 
\bottomrule
\end{tabular}
\label{tab:samples:bsm}
\end{table}

The effect of multiple interactions in the same and neighbouring bunch crossings (\pileup) was modelled by overlaying each simulated hard-scattering event with inelastic proton--proton ($pp$) events generated by \PYTHIA[8.186] using the \NNPDF[2.3lo] set of PDFs
and the A3 tune~\cite{ATL-PHYS-PUB-2016-017}. The generated Higgs boson events from Table~\ref{tab:samples:nominal_mc}  were passed through a \GEANT~\cite{Agostinelli:2002hh}
simulation of the ATLAS detector~\cite{SOFT-2010-01}. The remaining QCD diphoton events and \BSM\ samples are passed
through a fast simulation of the ATLAS detector to reduce the amount of CPU time
needed to process the large numbers of generated events. All simulated events are reconstructed with the same software as used for the data~\cite{ATL-SOFT-PUB-2021-001}.

% End of text imported from the .//sections/monte.tex input file

\section{Object and event selection}
\label{sec:data}

% The next lines are included from the .//sections/data.tex input file

\subsection{Object definitions}
 
Events in this analysis are selected using the following procedure. First, reconstructed photon candidates are required to satisfy a set of preselection criteria. The two highest-\pt preselected photons along with the reconstructed vertex information in the event, are used as inputs to a neural-network algorithm trained on simulated events to determine the correct primary vertex, which is named the \enquote{diphoton vertex}~\cite{HIGG-2013-08}. The selected primary vertex is used to compute the properties of objects in the event. Finally, to be selected the photons are required to satisfy isolation and additional identification criteria. In addition to photons, jets (including $b$-jets), muons, electrons, and the amount of missing transverse momentum, \met, are used in the analysis to define model-independent signal regions. The specific requirements for objects used in this analysis are as follows.

Photons must satisfy $\pt > 22$~\GeV\ and $|\eta|<2.37$, excluding the transition region $1.37<|\eta|<1.52$ between the barrel and endcap EM calorimeters.
Photon candidates are separated from jet backgrounds using a \emph{tight} identification
criterion based on calorimeter shower shape variables \cite{EGAM-2018-01}. The identification efficiency for reconstructed photons ranges from 84\% at $\pt = 25$~\GeV\ to 94\% at $\pt > 100$~\GeV.
The final selection of photons includes both calorimeter- and track-based isolation requirements to further suppress jets misidentified
as photons. The calorimeter isolation variable is defined as the energy in the EM calorimeter
within a cone of size $\Delta R=0.2$ around the photon candidate, excluding the energy in a fixed-size window that contains the photon shower; a correction is made for photon energy leakage from this window~\cite{EGAM-2018-01}. Contributions from \pileup and the underlying event are subtracted~\cite{Cacciari:2008gn,Cacciari:2009dp,PERF-2013-04,STDM-2010-08,EGAM-2018-01}.
The calorimeter-based isolation must be less than 6.5\% of each photon candidate's transverse energy.
The track-based isolation variable is defined as the scalar sum of the transverse momenta of tracks within
a $\Delta R = 0.2$ cone around the photon candidate. The tracks used in the isolation variable
are restricted to those with $\pt>1$~\GeV\ that are associated with the selected diphoton vertex and
not with a photon conversion vertex~\cite{EGAM-2018-01}.
The track isolation must be less than 5\% of each photon candidate's transverse energy.

Electron candidates must have $\pt>10$~\GeV\ and $|\eta|<2.47$, excluding the EM calorimeter transition region of $1.37<|\eta|<1.52$,  and must satisfy a \emph{medium} selection based on a likelihood discriminant using
calorimeter shower shapes and track parameters~\cite{EGAM-2018-01}.
Isolation criteria based on calorimeter- and track-based
information are applied to electrons. The reconstructed track matched to the electron candidate must be consistent with the diphoton vertex by requiring its longitudinal impact parameter $z_0$ relative to the vertex to satisfy $|z_0 \sin\theta|<0.5$~mm. In addition, the electron's transverse impact parameter $d_0$ with respect to the
beam-spot must satisfy $|d_0|/\sigma_{d_0} < 5$, where $\sigma_{d_0}$ is the uncertainty in $d_0$.

Muon candidates are required to have $\pt>10$~\GeV\ and $|\eta|<2.7$, and must satisfy the {\itshape medium}
identification requirements \cite{PERF-2015-10}.
Muons are required to satisfy calorimeter- and track-based isolation requirements that are
95\%--97\% efficient for muons with $\pt\in[10,60]$~\GeV\ and 99\% efficient for $\pt>60$~\GeV.
Muon tracks must satisfy $|z_0 \sin\theta|<0.5$~mm and $|d_0|/\sigma_{d_0}<3$.
 
Jets are reconstructed using a particle-flow algorithm~\cite{PERF-2015-09}. It improves the energy resolution by applying the anti-$k_t$ algorithm~\cite{Cacciari:2008gp,Fastjet} with radius parameter $R = 0.4$ to noise-suppressed positive-energy topological clusters~\cite{PERF-2014-07} in the calorimeter after removing energy deposits associated with primary-vertex-matched tracks, and including the track momenta in the clustering instead.
Jets must have $\pt>25$~\GeV\ and $|y|<4.4$.
To suppress jets from \pileup, a jet-vertex-tagger (JVT) multivariate discriminant~\cite{PERF-2014-03} is applied to jets with $\pt < 60$~\GeV\ and $|\eta|<2.4$; in the range $|\eta| > 2.5$, a `forward' version of the JVT~\cite{ATL-PHYS-PUB-2019-026} is applied to jets with $\pt < 120$~\GeV.\footnote{$\eta$ is used to select jets when applying the JVT and the $b$-tagging requirements, as the way these selections are defined is based on the jet position in the detector.}
Jets with $|\eta|<2.5$ containing $b$-hadrons are identified using the DL1r $b$-tagging algorithm and its 60\%, 70\%, 77\% and 85\% efficiency working points, with the outputs combined into a pseudo-continuous $b$-tagging score~\cite{FTAG-2018-01}.

Hadronically decaying $\tau$-leptons are used in this search. They are required to have $\pt>20$~\GeV\ and $|\eta|<2.7$. Furthermore, they must have either one or three charged tracks (\enquote{prongs}) with a charge sum of $\pm$ 1 in units of the elementary charge, and must satisfy the \emph{medium} working point~\cite{ATL-PHYS-PUB-2019-033}.
 
An overlap removal procedure is performed in order to avoid double-counting of objects.
First, electrons overlapping with any of the two selected photons ($\Delta R<0.4$) are removed.
Jets overlapping with the selected photons ($\Delta R<0.4$) and electrons ($\Delta R<0.2$) are removed. Electrons
overlapping with the remaining jets ($\Delta R<0.4$) are removed to match the requirements imposed when
measuring isolated-electron efficiencies.
Finally, muons overlapping with photons or jets ($\Delta R<0.4$) are removed.
 
The missing transverse momentum is defined as the negative vector sum of the
transverse momenta of the selected photons, electrons, muons, and jets, as well as the
transverse momenta of remaining low-\pt particles, estimated using tracks associated with the diphoton
vertex but not with any selected object~\cite{Aaboud:2018tkc}.

\subsection{Event preselection}
The two highest-\pt\ preselected photon candidates are required to satisfy the {\itshape tight} identification criteria and the isolation selection described above.
Finally, the leading and subleading photon candidates are required to satisfy $\pt^{\gamma1} > 35$~\GeV,  $\pt^{\gamma1} / m_{\gamma\gamma}> 0.35$, and $\pt^{\gamma2} > 25$~\GeV, $\pt^{\gamma2} / m_{\gamma\gamma}> 0.25$, respectively.
Events that fail the tight identification or the isolation selection but pass the loose identification are used as a control sample for background estimation and modelling purposes.
 
The trigger, object, and event selection described above are used to define the events that are selected for further
analysis. In total, about 1.2~million events are selected in this data set with a diphoton invariant mass between
105 and 160~\GeV. 

% End of text imported from the .//sections/data.tex input file

\section{Signal regions \label{sec:region}}

% The next lines are included from the .//sections/signalRegions.tex input file
Preselected events are assigned to  signal regions defined by the presence of additional reconstructed objects.  These signal regions are designed with simple, inclusive selections, and the specific selections were not optimized for any particular \BSM\ processes.
Since the selection criteria are not designed to be orthogonal, selected events can be assigned to multiple signal regions.
The overlap and correlations of these signal regions are discussed in Section~\ref{sec:result}. Based on the requirement of there being additional objects in the event, the signal regions can be broadly classified to six groups: events with heavy-flavour jets; events with high jet multiplicity or a large scalar sum of jet transverse momenta (\HT); events with large \met; events with leptons; events with additional photons; and events with top quarks. As is discussed in Section~\ref{sec:result},
each signal region, defined through selection criteria for detector-level objects, has a counterpart defined through equivalent criteria at the particle level,
which enables the interpretation of the search results with particle-level simulation samples.
These signal regions and their definitions are summarized in Table~\ref{tab:regsummary} and described as follows.
 
Two \enquote{heavy-flavour} signal regions are defined. One requires the presence of at least three $b$-jets tagged at the 85\% efficiency working point (WP), and the other requires the presence of at least four $b$-jets tagged at the same working point.
 
Six high-jet-activity signal regions are defined. Three of them are defined with inclusive jet-multiplicity requirements, using central jets with $|\eta| < 2.5$. The thresholds for the number of jets are four, six, and eight for these regions. Another three regions are defined by the \HT. The inclusive \HT\ thresholds for these three regions are 500~\GeV, 1000~\GeV, and 1500~\GeV.
 
Three high-\met\ signal regions are defined using \met\ thresholds of 100~\GeV, 200~\GeV, and 300~\GeV.
 
Six leptonic signal regions are defined. The inclusive leptonic signal region (${\ge}1\ell$) requires the presence of at least one electron or muon in the event. Three dilepton signal regions are defined with electrons and muons: the inclusive dilepton signal region ($2\ell$) requires the presence of exactly two leptons; the dilepton $Z$-veto ($2\ell$--$\cancel{Z}$) region is a subset of the $2\ell$ region, and it explicitly rejects events where the two leptons have the same flavour and their invariant mass is within 10~\GeV\ of the PDG value of the $Z$ boson mass; the same-sign dilepton (SS-$2\ell$) region is also a subset of the $2\ell$ region and requires the electric charges of the two leptons to be the same; The multilepton signal region (${\geq}3\ell$) requires the presence of at least three leptons. The $\tau$-lepton signal region requires the presence of at least two $\tau$-lepton candidates.
 
Two triphoton signal regions are defined. Both require the presence of at least one additional photon in the preselected diphoton events. In the first photon signal region, the diphoton system that is considered as the Higgs boson candidate is built from the two leading photons in \pt. The second photon region assumes the subleading and third-leading photons come from the Higgs boson decay. While the two photon signal regions mostly have the same definitions of the event selection criteria, the \mgg\ observable is constructed differently. The 1$\gamma$--$m_{\gamma\gamma}^{12}$ region targets exotic processes where the Higgs boson is accompanied by a photon with moderate \pt, while the 1$\gamma$--$m_{\gamma\gamma}^{23}$ region targets exotic processes in which a new state that is much more massive than the Higgs boson decays into a photon and a Higgs boson.
 
Three \enquote{top} signal regions are introduced. One inclusive top signal region ($\ell b$) requires the presence of at least one electron or one muon and at least one $b$-jet tagged at the 70\% efficiency working point. One exclusive semileptonic top-quark signal region ($t_\text{lep}$) requires the presence of exactly one lepton, and one $b$-jet tagged at the 70\% efficiency working point. One exclusive hadronic top-quark signal region ($t_\text{had}$) requires the presence of exactly three jets, one of which must be a $b$-jet tagged at the 70\% efficiency working point. The hadronic top-quark signal region also vetoes events with at least one electron or one muon. To further suppress diphoton-plus-multijet background, a boosted decision tree (BDT) is used in top-quark reconstruction~\cite{HIGG-2019-01} performed with the three jets in the event, and the resulting BDT score is required to be greater than 0.9.

\begin{table}[bh!]
\begin{center}
\caption{Signal region definitions at the detector level and particle level. Particle-level object definitions are detailed in Section~\ref{sec:result:interpretation}. The \HT\ variable is defined as the scalar sum of the transverse momenta of jets that have $|\eta|< 2.5$. WP stands for efficiency working point.
At the detector level, the $\tau$-lepton identification is optimized for hadronically decaying $\tau$-leptons.
The symbols $\gamma_1$, $\gamma_2$, and $\gamma_3$ denote the leading, subleading, and third-leading photons ordered in \pt, respectively. The BDT$_\text{top}$ variable is a boosted decision tree (BDT) score used to identify triplets of jets consistent with the hadronic-decay final states of top quarks, which was used in Ref.~\cite{HIGG-2020-16}.}
\label{tab:regsummary}
\begin{tabular}{m{2cm}m{2.5cm}m{5cm}m{5cm}}
\hline
Target & Region     &Detector level & Particle level \\
\hline
\multirow{2}{2cm}{Heavy flavour} & $\geq$3$b$   & $n^\mathrm{}_{\bjet} \ge 3$, 85\% WP & $n^\mathrm{}_{\bjet} \ge 3$ \\
& $\geq$4$b$ & $n^\mathrm{}_{\bjet} \ge 4$, 85\% WP & $n^\mathrm{}_{\bjet}~\ge$ 4\\
\hline
\multirow{6}{2cm}{High jet activity} & $\geq$4j   & $n^\mathrm{}_\mathrm{jet}~\ge$ 4, $|\eta_\mathrm{jet}|~<$ 2.5 & $n^\mathrm{}_\mathrm{jet}~\ge$ 4, $|\eta_\mathrm{jet}|~<$ 2.5 \\
& $\geq$6j  &$n^\mathrm{}_\mathrm{jet}~\ge$ 6, $|\eta_\mathrm{jet}|~<$ 2.5 & $n^\mathrm{}_\mathrm{jet}~\ge$ 6, $|\eta_\mathrm{jet}|~<$ 2.5 \\
& $\geq$8j  &$n^\mathrm{}_\mathrm{jet}~\ge$ 8, $|\eta_\mathrm{jet}|~<$ 2.5 & $n^\mathrm{}_\mathrm{jet}~\ge$ 8, $|\eta_\mathrm{jet}|~<$ 2.5 \\
&$\HT>500$~\GeV  &$\HT > 500$~\GeV&$\HT > 500$~\GeV\\
&$\HT>1000$~\GeV &$\HT > 1000$~\GeV &$\HT > 1000$~\GeV\\
&$\HT>1500$~\GeV &$\HT > 1500$~\GeV &$\HT  > 1500$~\GeV\\
\hline
 
\multirow{3}{2cm}{\met} & \met$>$100 \GeV  & $\met > 100$~\GeV & \trumet $> 100$~\GeV\\
& \met$>$200 \GeV  & $\met > 200$~\GeV & \trumet $> 200$~\GeV \\
& \met$>$300 \GeV  & $\met > 300$~\GeV & \trumet $> 300$~\GeV \\
\hline
\multirow{3}{2cm}{Top} &  $\ell b$ & $n_{\ell=e,\mu}\ge1$, $n^\mathrm{}_{\bjet} \ge 1$, 70\%~WP & $n_{\ell=e,\mu}\ge1$, $n^\mathrm{}_{\bjet} \ge 1$ \\
&$t_\text{lep}$ &  $n_{\ell=e,\mu}=1$, $n_\mathrm{jet}=n^\mathrm{}_{\bjet}=1$, 70\%~WP & $n_{\ell=e,\mu}=1$, $n_\mathrm{jet}=n^\mathrm{}_{\bjet}=$ 1  \\
&$t_\text{had}$ & $n_{\ell=e,\mu}=0$, $n_\mathrm{jet}=3$, $n^\mathrm{}_{\bjet}= 1$, 70\%~WP, BDT$_\text{top}>0.9$ & $n_{\ell=e,\mu}=0$, $n_\mathrm{jet}=3$, $n^\mathrm{}_{\bjet}= 1$ \\
\hline
 
\multirow{6}{2cm}{Lepton}& $\geq$1$\ell$ & $n_{\ell=e,\mu}\ge1$ & $n_{\ell=e,\mu}\ge1$  \\
& 2$\ell$ &  $ee,\mu\mu$, or $e\mu$ & $ee,\mu\mu$, or $e\mu$ \\
& 2$\ell$--$\cancel{Z}$  & $ee,\mu\mu,e\mu$; $|m_{\ell\ell}-m_Z| > 10$~\GeV\ for same-flavour leptons & $ee,\mu\mu,e\mu$; $|m_{\ell\ell}-m_Z| > 10$~\GeV\ for same-flavour leptons \\
& SS-2$\ell$  & $ee$, $\mu\mu$, or $e\mu$ with same charge & $ee$, $\mu\mu$, or $e\mu$ with same charge \\
& $\geq$3$\ell$& $n_{\ell=e,\mu}\ge3$  & $n_{\ell=e,\mu}\ge3$\\
& $\geq$2$\tau$   & $n_{\tau,\text{had}}\ge2$ 
& $n_{\tau}\ge2$ \\
 
\hline
 
\multirow{2}{2cm}{Photon} & 1$\gamma$--$m_{\gamma\gamma}^{12}$ & $n_{\gamma}\ge3$, $m_{\gamma\gamma}$ defined with $\gamma_1$,$\gamma_2$ & $n_{\gamma}\ge3$, $m_{\gamma\gamma}$ defined with $\gamma_1$,$\gamma_2$\\
& 1$\gamma$--$m_{\gamma\gamma}^{23}$  &$n_{\gamma}\ge3$, $m_{\gamma\gamma}$ defined with $\gamma_2$,$\gamma_3$ & $n_{\gamma}\ge3$, $m_{\gamma\gamma}$ defined with 
$\gamma_2$,$\gamma_3$\\
\hline
\end{tabular}
\end{center}
\end{table}
 
\FloatBarrier

% End of text imported from the .//sections/signalRegions.tex input file

\section{Signal and background modelling}
\label{sec:modeling}

% The next lines are included from the .//sections/modeling.tex input file
This search uses the invariant mass of the diphoton system (\mgg) as the discriminating variable. The potential BSM signal is a narrow resonance in \mgg, and the background consists of a resonant component and a continuum component. The resonant component arises from the SM Higgs boson production processes, which include gluon--gluon fusion (\ggF) and vector-boson fusion (VBF), and \VH, \ttH, \tH, and \bbH\ production.
 
The continuum component arises from the production of two initial- or final-state photons, or from the misidentification of jets as either one or both of the photons selected at the detector level.
In signal regions where lepton and/or \bjet requirements are applied, these three types of continuum background processes also involve the production of electroweak bosons or top quarks.
 
The \mgg\ distribution in each signal region is described by a probability density function (pdf) where the signal and background components are modelled by analytic functions of \mgg and are normalized to their expected yields. As in previous ATLAS $H \rightarrow \gamma\gamma$ measurements~\cite{HIGG-2020-16,HIGG-2019-13}, the analytic functions are defined in the range of $105~\GeV < \mgg < 160$~\GeV.
In the statistical interpretation of the search results, the \mgg\ pdfs corresponding to the signal-plus-background and background-only hypotheses are used to fit the observed \mgg\ distributions.
 
The event yield in the multilepton signal region is not sufficient for performing a fit to the \mgg\ distribution. In this case, an event-count analysis is performed, as is described in Section~\ref{sec:SRmultilepton}.
 
\subsection{Modelling of BSM and SM Higgs boson production}
 
The modelling of the \mgg\ distribution and its associated systematic uncertainties is done in a consistent way for the BSM and SM Higgs boson production processes. Their \mgg\ distributions are both described by the same double-sided Crystal Ball (DSCB) function~\cite{Oreglia:1980cs,HIGG-2014-04}, consisting of a Gaussian distribution in the region around the peak position, continued by power-law tails at lower and higher \mgg\ values. The potential bias in the estimated signal yield due to an intrinsic shape difference between the DSCB function and signal \mgg\ distribution is found to be much smaller than the MC statistical uncertainty of the signal yield and therefore is negligible~\cite{HIGG-2016-21}. The parameters of the DSCB function in each signal region are obtained by a fit to a mixture of the \ggF, VBF, \VH, \ttH, \tH, and \bbH\ samples described in Section~\ref{sec:monte}, which are normalized according to their SM cross sections. A shift of $0.09\,\GeV$ is applied to the position of the signal peak to account for the difference between the reference Higgs boson mass used in this analysis ($m_H = 125.09\,\GeV$)~\cite{HIGG-2014-14} and the mass for which the samples were generated ($m_H = 125\,\GeV$). The uncertainty in the \mgg\ modelling is discussed in Section~\ref{sec:syst:mgg}. There is a modest variation in the DSCB shape parameters between various SM and BSM Higgs boson production samples. However, the impact of this variation on the fitted number of signal events is much smaller than the statistical uncertainty of the fitted signal yield.
 
SM Higgs boson production is the resonant background to signals arising from BSM production. The cross sections of the resonant backgrounds are assumed to be the same as the theoretical predictions, and their contributions to various signal regions are evaluated using the MC samples described in Section~\ref{sec:monte}. The recent Higgs boson measurements from the ATLAS and CMS experiments have shown that the data are consistent with the theory predictions within statistical and systematic uncertainties, which justifies the fixing of expected resonant background yields to their SM predictions.
Potential interference effects between the SM and BSM Higgs boson production processes are not considered in the MC simulation. If the interference effect results in a significant deviation in the production rate from the SM prediction, this search would detect it as a signal.

\subsection{Continuum background modelling}
\label{sec:continuumbkg}
The \mgg\ distribution of the continuum background is fully determined from a fit to data. The only systematic uncertainty associated with this method arises from the intrinsic difference between the true continuum background \mgg\ shape and the shape of the chosen analytic pdf.
 
The modelling of the continuum background follows the same process used in Ref.~\cite{HIGG-2016-21}, which involves two main steps: first, a background \mgg\ template is constructed from either the MC diphoton continuum background or a data control sample where at least one of the two leading photons fails the photon identification or isolation requirement. Between these two samples, the one with a higher number of events is selected to construct the \mgg\ template. Second, a background function is selected from a number of candidate functions, using a procedure known as the \emph{spurious-signal test}~\cite{HIGG-2016-21}, with the goal of identifying an analytic function that is flexible enough to fit the \mgg\ distribution in data and which results in a small potential bias compared to the background statistical uncertainty.
 
Several families of analytic functions are tested as candidates to model the \mgg\ distribution for each signal region. They include power-law functions, Bernstein polynomials, and exponential functions of a polynomial. These functional forms can include up to five free parameters. The selected functional forms are mostly power-law and exponential functions with one free parameter.
The coefficients of background functions are considered independent across signal regions, regardless of the functions chosen. In all cases, they are treated as free parameters in the fit to data.

\subsection{Modelling of the background in the multilepton signal region}
\label{sec:SRmultilepton}
The expected continuum background event yield in the multilepton (${\geq}3\ell$) signal region is so small that the \mgg\ distribution may not be fitted with an analytic function. In this case, the event count in the \mgg\ range from 123~\GeV\ to 127~\GeV\ is used for statistical interpretation. The expected background yield in this mass range is extrapolated from a control region in two steps. This control region is referred to as the \emph{non-tight-isolated} region and is defined in the same way as the multilepton signal region, except that at least one of the two leading photons must fail to meet the photon quality criteria detailed in Section~\ref{sec:data}. The expected background yield in the diphoton mass sidebands of the multilepton signal region is scaled from the event yield in the non-tight-isolated diphoton mass sidebands. The ratio of the event yield in the non-tight-isolated control region to that in the tight-isolated signal region is checked in the other leptonic signal regions and is found to be consistent, which indicates that the scale factor derived from other leptonic signal regions can be applied to the multilepton signal region.  The variations in the scale factor between various leptonic signal regions are considered as systematic uncertainties of the scale factor. Similarly, the expected background yield in the mass window of $123~\GeV < \mgg < 127$~\GeV\ is computed from the expected background yield in the diphoton mass sidebands, using scale factors evaluated in the non-tight-isolated control regions defined for various leptonic signal regions. These scale factors are consistent and their variations are considered as systematic uncertainties of the extrapolation. The expected continuum background yield in the multilepton signal region is found to be $0.25^{+0.59}_{-0.25}$ events, where the lower bound of the uncertainty is truncated so that the yield remains positive.
The expected resonant background yield in this region is predicted by the MC simulation.

% End of text imported from the .//sections/modeling.tex input file

\section{Systematic uncertainties}
\label{sec:systematics}

% The next lines are included from the .//sections/systematics.tex input file
Systematic uncertainties considered in this analysis can be grouped into three categories: uncertainties in the modelling of the \mgg\ distribution for the BSM and SM Higgs boson production processes, uncertainties in the modelling of the continuum background \mgg\ distribution, and uncertainties in the expected resonant background yields in each signal region arising from either experimental or theoretical sources.  These systematic uncertainties are incorporated into the analysis likelihood models as nuisance parameters.
 
\subsection{Systematic uncertainties in modelling \mgg\ for SM and BSM Higgs boson production}
\label{sec:syst:mgg}
 
In general, the \mgg\ modelling systematic uncertainties for a given region depend on the BSM signal considered. However, the signal shape variations associated with the properties of a specific BSM model are negligible relative to those associated with the experimental effects, and only the latter are therefore considered and estimated from the SM Higgs production processes.
The extraction of a potential signal component from the data \mgg\ distribution is subject to systematic uncertainties in the diphoton mass resolution and scale. While the diphoton mass resolution and scale uncertainties are also model-dependent, their variations are typically limited. The likelihood model considers only signal systematic uncertainties related to the diphoton mass scale and resolution, and does not include any systematic uncertainty for the signal yield, which is the parameter of interest in the search.
 
The diphoton mass scale uncertainty includes a contribution from the photon energy scale uncertainty, which is typically less than 1\% of the diphoton mass, and also a systematic component of 240 MeV, corresponding to the measurement uncertainty of the Higgs boson mass from the combination of the ATLAS and CMS data~\cite{HIGG-2014-14}. The diphoton mass resolution uncertainty arises from the uncertainty of the photon energy resolution and is about 3\%--15\% relative to the reconstructed width of the diphoton resonance.
The estimation and implementation of the photon energy scale and resolution uncertainties follow the procedure outlined in Ref.~\cite{EGAM-2018-01}.

\subsection{Uncertainty in the continuum background modelling}
The uncertainty in the modelling arises from an intrinsic shape difference between the continuum background \mgg\ distribution and the chosen analytic function. The size of this uncertainty is estimated using the \emph{spurious-signal} test mentioned in Section~\ref{sec:continuumbkg} and is always smaller than 20\% of the statistical uncertainty of the continuum background.
 
\subsection{Uncertainties in resonant background yields}
The definition of the signal regions includes selections based on photons, leptons, jets (\bjets), and missing transverse momentum. As such, the expected resonant background yield is subject to experimental systematic uncertainties in the reconstruction and identification efficiencies of these objects.
The typical level of systematic uncertainty for object reconstruction and identification is less than 1\% for photons~\cite{EGAM-2018-01}, 1\%--4\% for electrons~\cite{EGAM-2018-01} and muons~\cite{MUON-2018-03}, 2\%--16\% for jets~\cite{JETM-2018-05} and $b$-jets~\cite{FTAG-2018-01}, and 2\% for $\tau$-leptons~\cite{ATLAS-CONF-2017-029}. Uncertainties in the theoretical modelling of SM Higgs boson production processes are estimated by varying the scale of QCD renormalization and factorization, by changing the choice of parton distribution function set, and by comparing the predicted events from different parton showering implementations for a given Higgs boson production process. The theoretical systematic uncertainties typically range from 10\% to 35\%~\cite{deFlorian:2016spz}. The uncertainty in the integrated luminosity of the data sample is 1.7\%~\cite{ATLAS-CONF-2019-021}, obtained using the LUCID-2 detector~\cite{LUCID2} for the primary luminosity measurements.
Other uncertainties in background event yields, such as the impact of \pileup on selection efficiencies~\cite{STDM-2015-05}, contribute $< 1\%$.

% End of text imported from the .//sections/systematics.tex input file

\section{Results}
\label{sec:result}

% The next lines are included from the .//sections/results.tex input file
 
\begin{figure}[htbp]
\centering
\subfloat[$\geq$$3b$]{\includegraphics[width=0.49\textwidth]{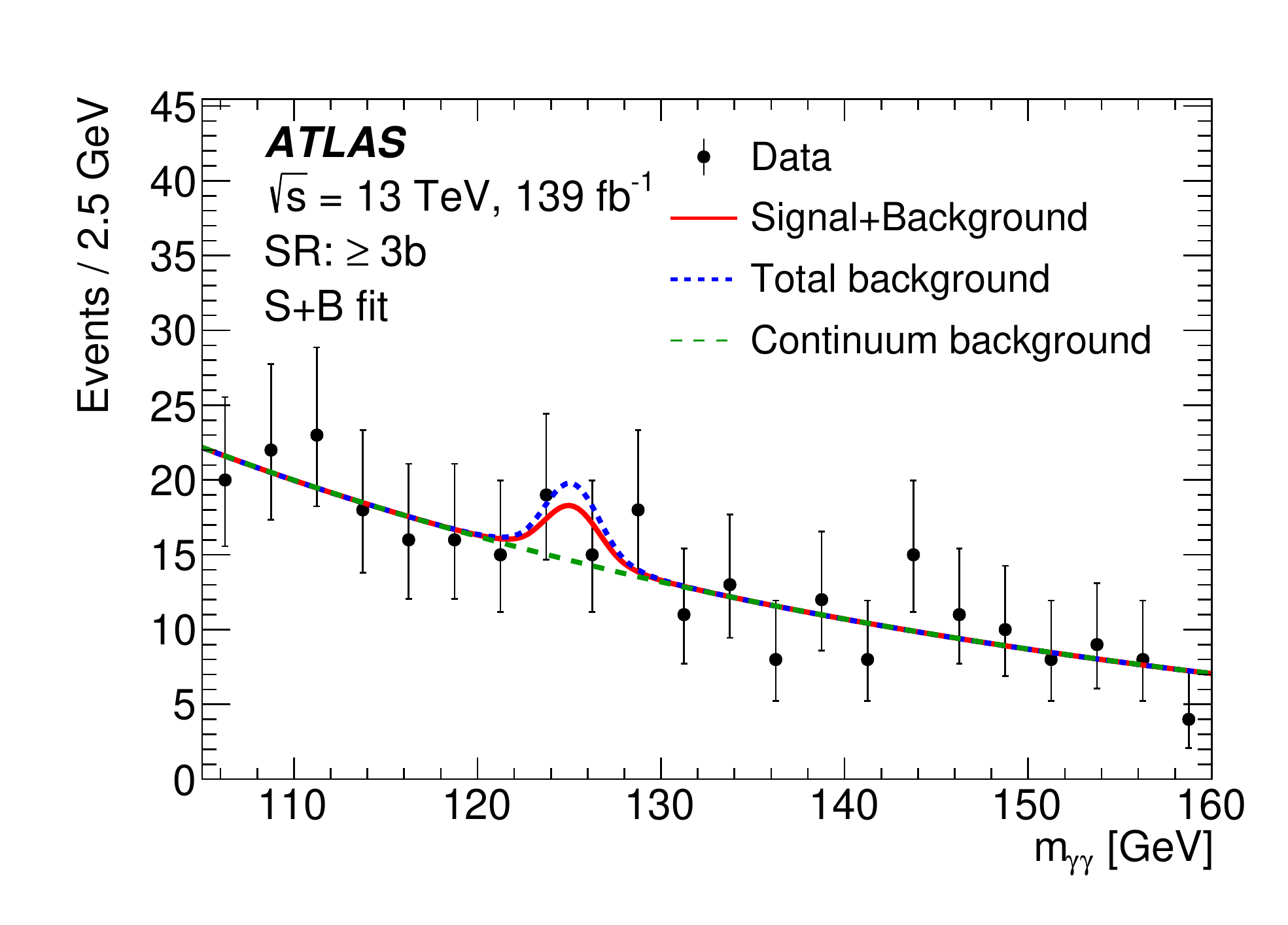}}
\subfloat[$\geq$$4b$]{\includegraphics[width=0.49\textwidth]{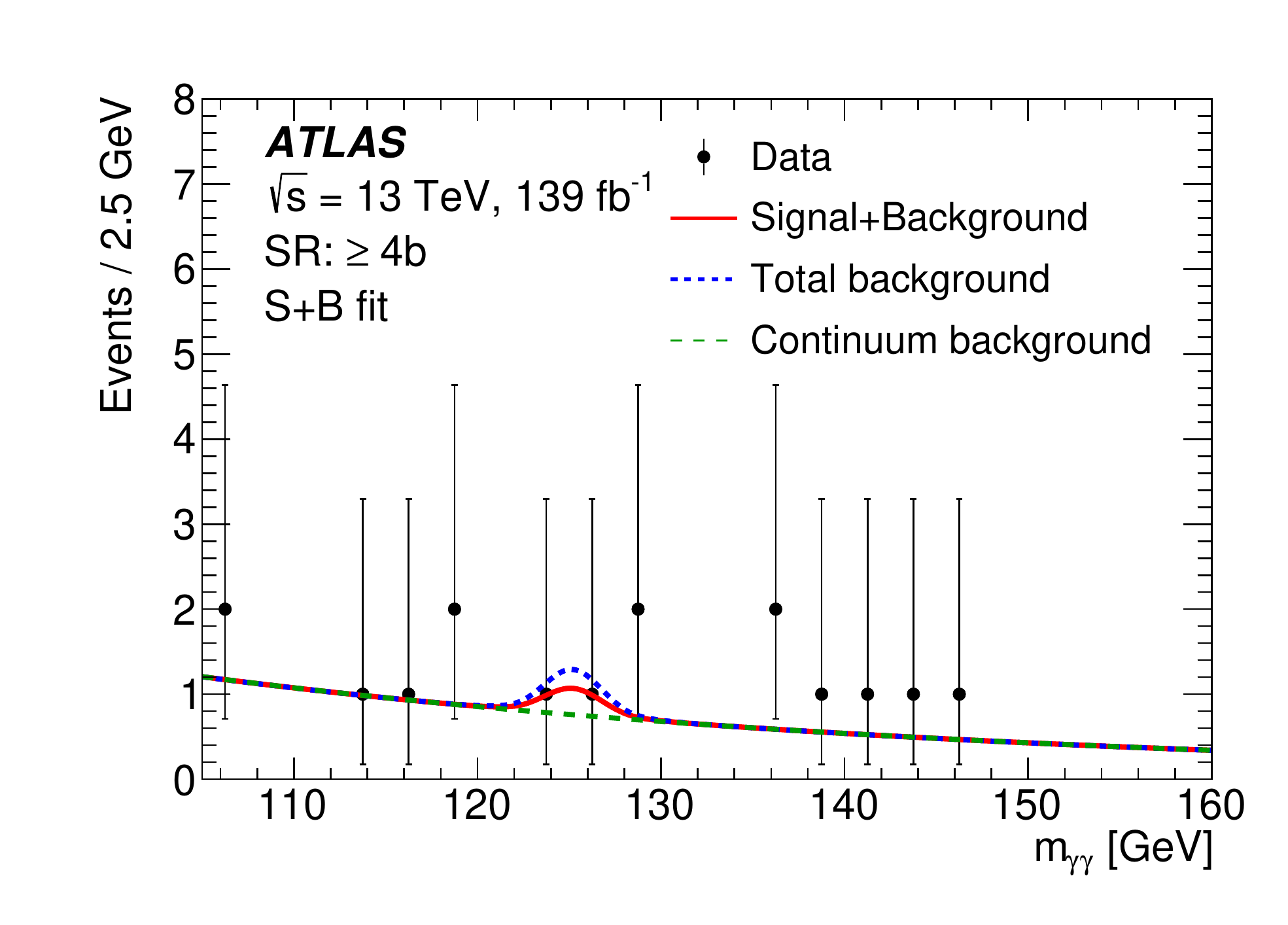}}
\caption{Diphoton mass distributions for heavy-flavour signal regions.
Data are shown together with the signal-plus-background fit. The overall fitted signal-plus-background pdf is shown as a solid red curve, the contributions from the resonant Higgs boson and continuum backgrounds are shown as a dashed blue curve, while the contributions only from the continuum backgrounds are shown as a dashed green curve. The difference between the solid red and the dashed blue curves indicates the fitted deviation from the SM expectation. }
\label{fig:unblind_fits_heavyflavor}
\end{figure}

\begin{figure}[htbp]
\centering
\subfloat[1$\gamma$--$m_{\gamma\gamma}^{12}$]{\includegraphics[width=0.49\textwidth]{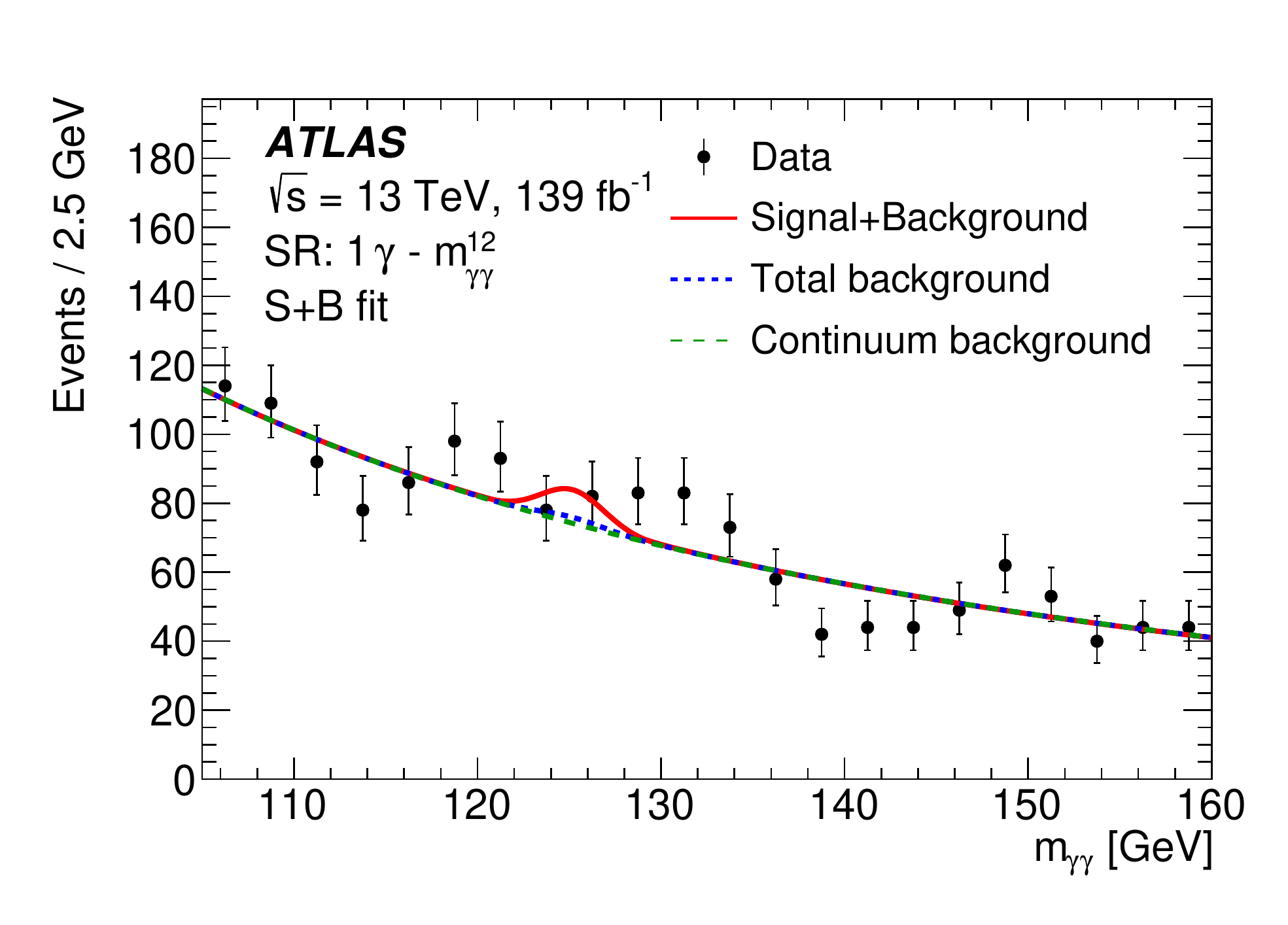}}
\subfloat[1$\gamma$--$m_{\gamma\gamma}^{23}$]{\includegraphics[width=0.49\textwidth]{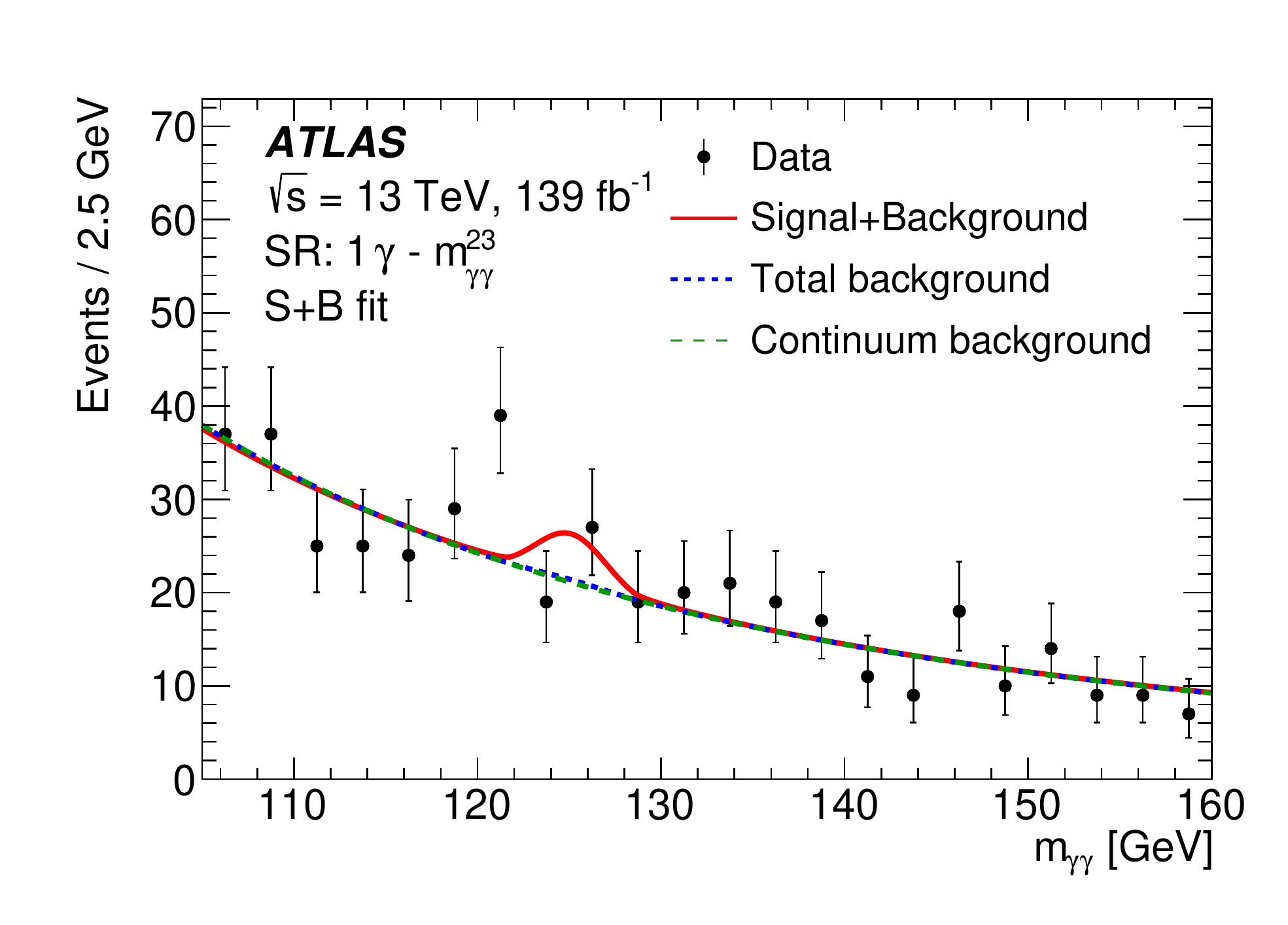}}\\
 
\caption{Diphoton mass distributions for photon signal regions.
Data are shown together with the signal-plus-background fit. The overall fitted signal-plus-background pdf is shown as a solid red curve, the contributions from the resonant Higgs boson and continuum backgrounds are shown as a dashed blue curve, while the contributions only from the continuum backgrounds are shown as a dashed green curve. The difference between the solid red and the dashed blue curves indicates the fitted deviation from the SM expectation. }
\label{fig:unblind_fits_photon}
\end{figure}

Figures~\ref{fig:unblind_fits_heavyflavor}--\ref{fig:unblind_fits_photon} show the diphoton mass \mgg\ distributions for all the signal regions considered in this paper, except the multilepton (${\geq}3\ell$) signal region. The result of a signal-plus-background fit as described in Section~\ref{sec:modeling} is shown for each signal region. The expected and observed event yields are determined from the signal-plus-background fit in each signal region and are summarized in Table~\ref{tab:res_bkgyields}. Also included in this table are the expected and observed event counts in the multilepton signal region. No event is found in the multilepton signal region, while the sum of the continuum and resonant background expectations is $0.27^{+0.59}_{-0.27}$ events . A $p_{0}$-value was evaluated for each of these signal regions, and the largest excess is in the \HT$>$1000 GeV region and is less than two standard deviations above the expectation. The observed significance~\cite{ATL-PHYS-PUB-2020-025} is also given in Table~\ref{tab:res_bkgyields}.

\begin{table}[h!]
\centering
\caption{Expected background and observed event yields in the diphoton mass range $123~\GeV < \mgg < 127$~\GeV.
The resonant Higgs boson production and continuum backgrounds are shown separately. The observed significance is also shown for all signal regions except the ${\ge}3\ell$ region, where less than one background event is expected and the observed number of events is zero.  Note that the significance is calculated using the full mass spectrum of 105 GeV $<$ \mgg\ $<$ 160 GeV, which may not be reproduced using yields reported in this table.}
\label{tab:res_bkgyields}
\scriptsize
\begin{tabular}{lccccc}
\toprule
\multirow{2}{1em}{SR}& \multicolumn{3}{c}{Expected Background} & \multicolumn{2}{c}{Observed} \\
& Resonant Higgs  & Continuum & Total Background & Observed Yield & Excess Significance [$\sigma$]\\
\hline
\multicolumn{6}{c}{Heavy flavour}\\
\hline
$\geq$3$b$ 						& 6.47 			& 23.4 				& 29.9 				& \phantom{0}30 	& --0.3 \\
$\geq$4$b$ 						& 0.69			& 1.22 		& 1.91 		& \phantom{00}1 	& --0.2 \\
\hline
\multicolumn{6}{c}{High jet activity}\\\hline
$\geq$4j 							& 85.2	& 1330	& 1420	& 1404\phantom{0} 	& --0.3 \\
$\geq$6j 							& 16.4	& 104		& 121 		& 105 			& --1.3 \\
$\geq$8j 							& 2.44 			& 6.37 		& 8.81 		& \phantom{00}6 	& --0.9 \\
\HT$>$\,500 \GeV					& 23.9	& 297		& 321	& 310			& --0.6 \\
\HT$>$\,1000 \GeV					& 1.85 			& 27 		& 28.8 				& \phantom{0}39 	& \phantom{--}1.8 \\
\HT$>$\,1500 \GeV					& 0.264	& 3.9 		& 4.17 		& \phantom{00}4	& \phantom{--}0.1 \\
\hline
\multicolumn{6}{c}{\met}\\\hline
\met$>$\,100 \GeV					& 29	& 171 		& 200 		& 212 			& \phantom{--}0.8 \\
\met$>$\,200 \GeV					& 4.51 			& 8.06 		& 12.6 				& \phantom{0}16 	& \phantom{--}0.9 \\
\met$>$\,300 \GeV					& 1.15 			& 1.85 		& 3 			& \phantom{00}5 	& \phantom{--}0.8 \\
\hline
\multicolumn{6}{c}{Top quark}\\ \hline
$\ell b$ 							& 14.9	& 27 		& 41.9 				& \phantom{0}34 	& --0.6 \\
$t_\text{lep}$ 						& 0.281	& 2.58 		& 2.86 		& \phantom{00}1 	& --0.7 \\
$t_\text{had}$ 						& 4.44 			& 96.3				& 101 		& 111 			& \phantom{--}1.7 \\
\hline
\multicolumn{6}{c}{Lepton}\\\hline
$\geq$1$\ell$ 						& 38.8	& 183		& 222		& 237 			& \phantom{--}1.4 \\
2$\ell$ 							& 4.24 			& 9.42		& 13.7 				& \phantom{0}10 	& --0.5 \\
2$\ell$--$\cancel{Z}$ 				& 1.95 			& 7.35 		& 9.3 		& \phantom{0}10 	& \phantom{--}0.7 \\
SS-2$\ell$ 						& 0.431	& 0.224 	& 0.655 	& \phantom{00}1 	& \phantom{--}0.2 \\
${\geq}3\ell$ 						& 0.02 			& 0.25 		& 0.27		& \phantom{00}0 	& -- \\
${\geq}2\tau$ 						& 0.256	& 0.875 	& 1.13 		& \phantom{00}2 	& \phantom{--}0.6 \\
\hline
\multicolumn{6}{c}{Photon}\\\hline
1$\gamma$--$m_{\gamma\gamma}^{12}$	& 2.33 			& 119 		& 121 		& 132 			& \phantom{--}0.7 \\
1$\gamma$--$m_{\gamma\gamma}^{23}$	& 0.436	& 32.8 				& 33.2 				& \phantom{0}42 				& \phantom{--}1.1 \\
\hline
\end{tabular}
\end{table}
 
\begin{figure}[htbp]
\centering
\subfloat[$\geq$4j]{\includegraphics[width=0.33\textwidth]{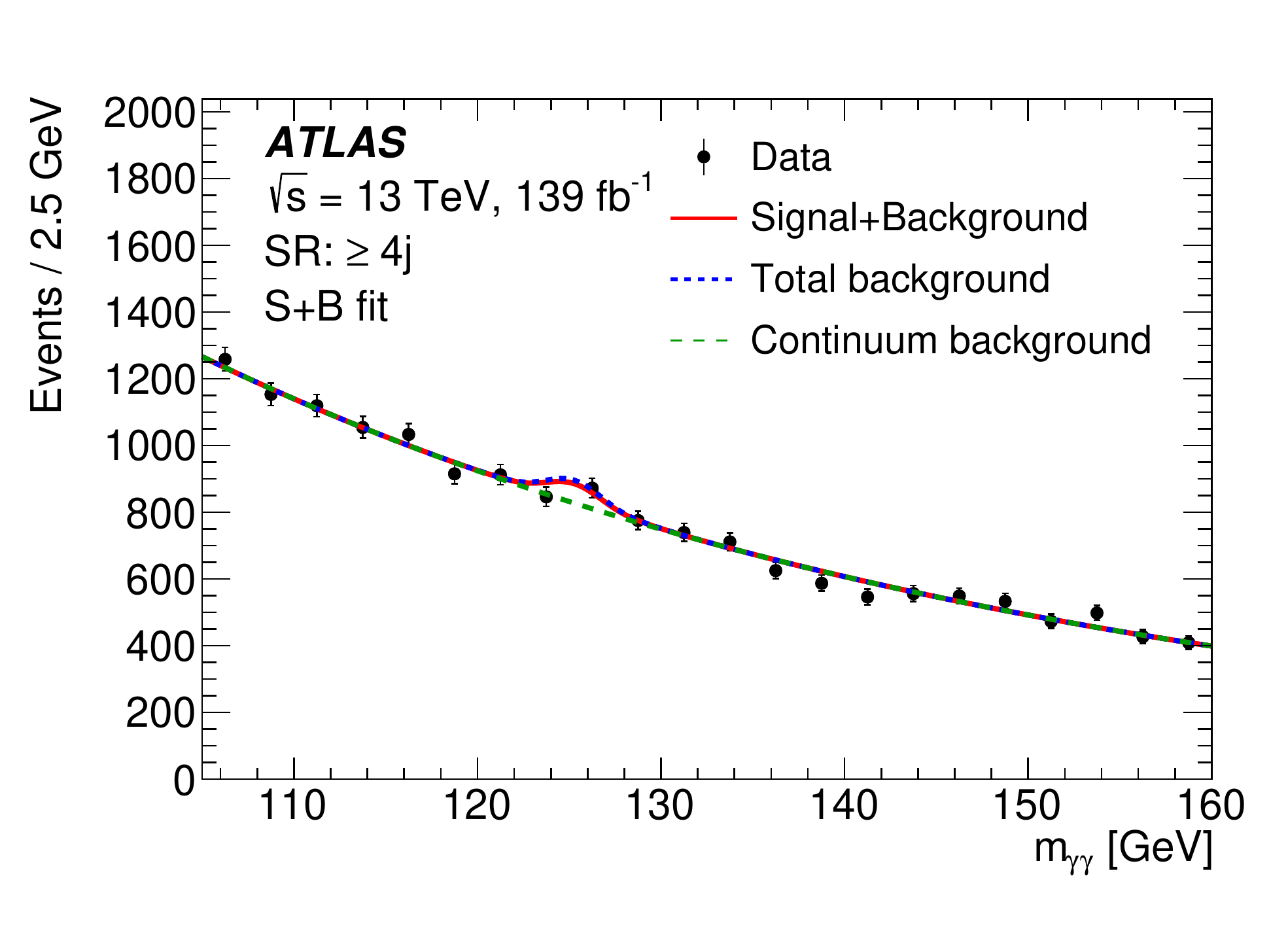}}
\subfloat[$\geq$6j]{\includegraphics[width=0.33\textwidth]{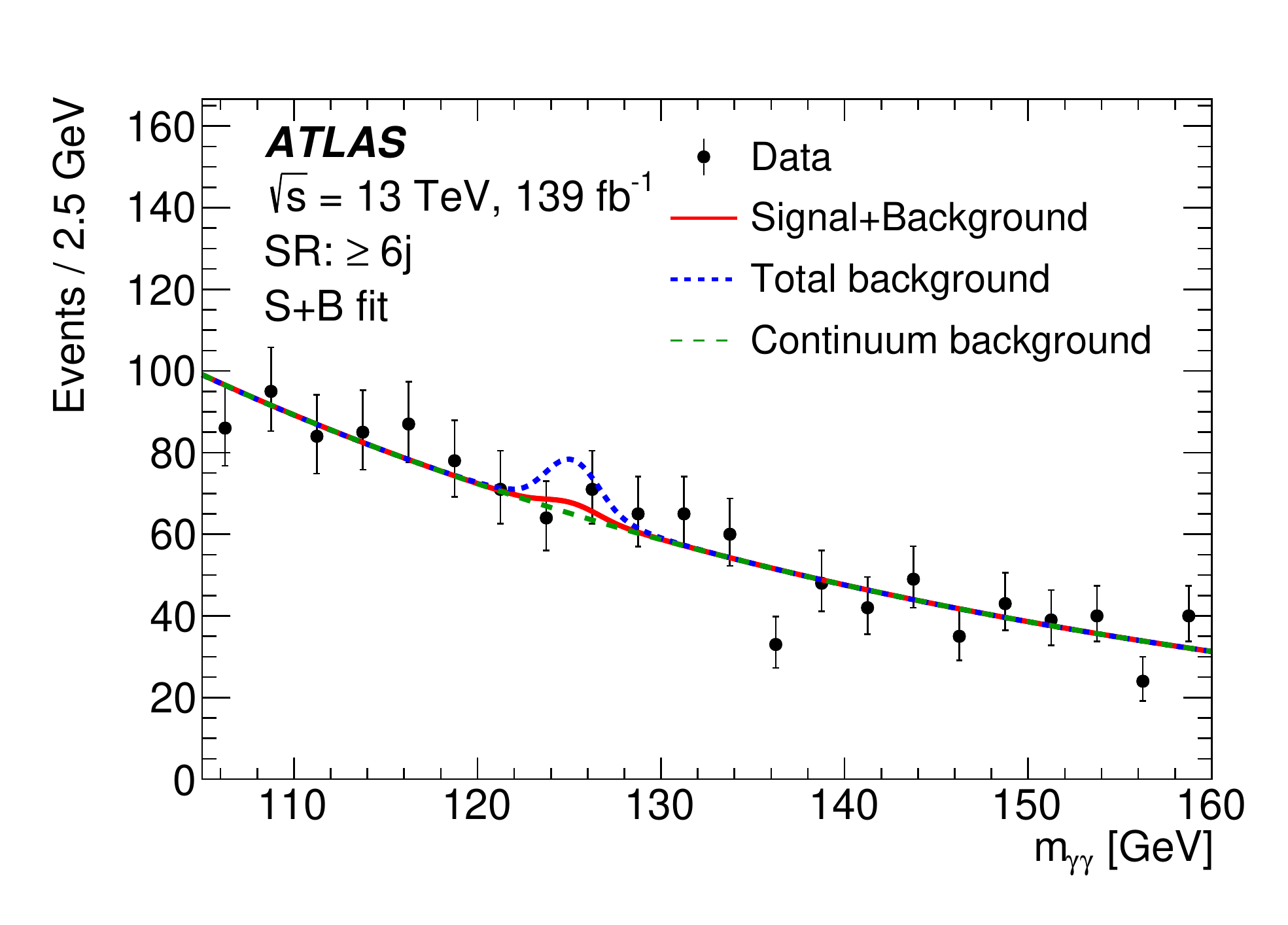}}
\subfloat[$\geq$8j]{\includegraphics[width=0.33\textwidth]{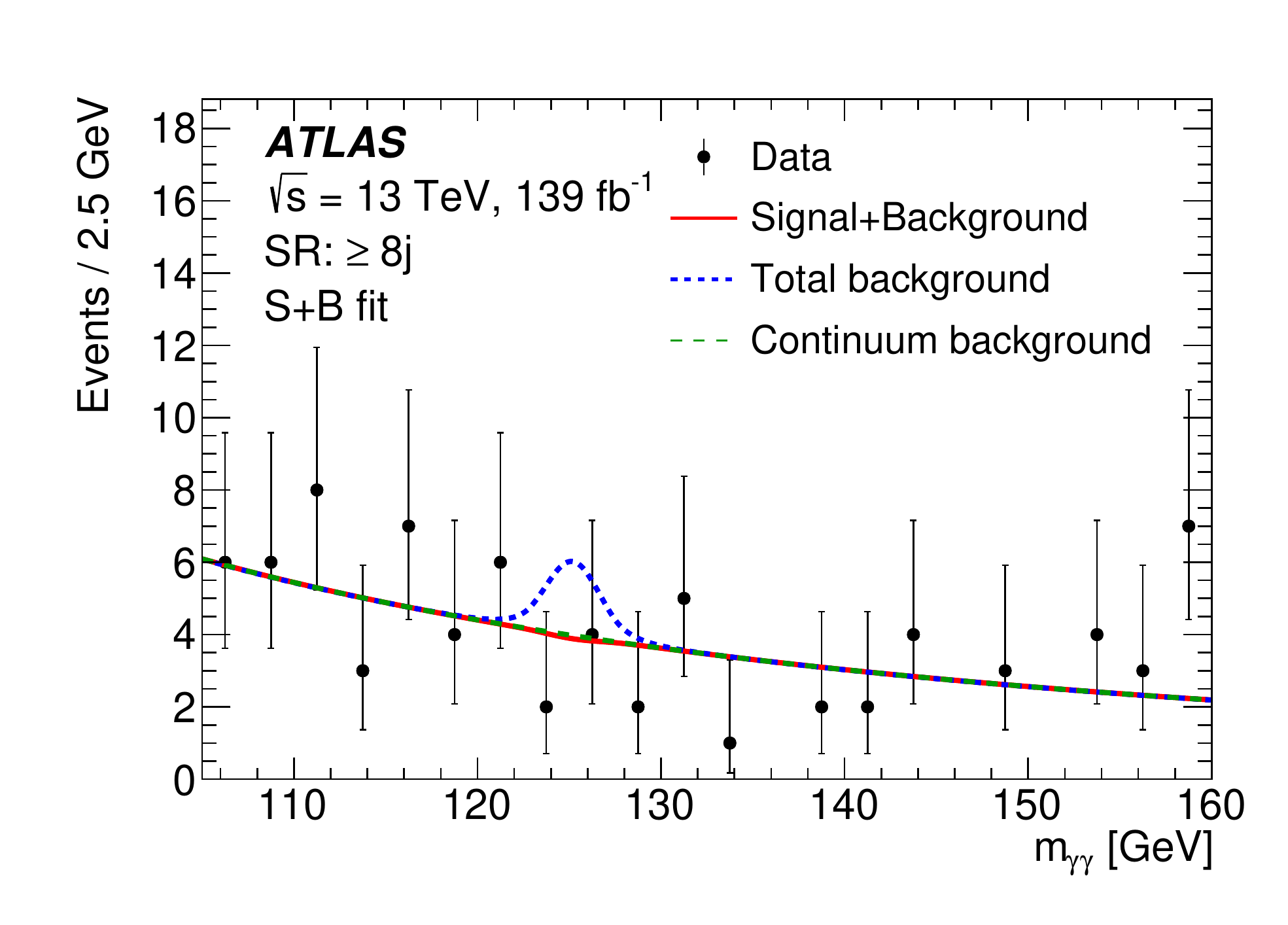}}\\
\subfloat[\HT$>$\,500 \GeV]{\includegraphics[width=0.33\textwidth]{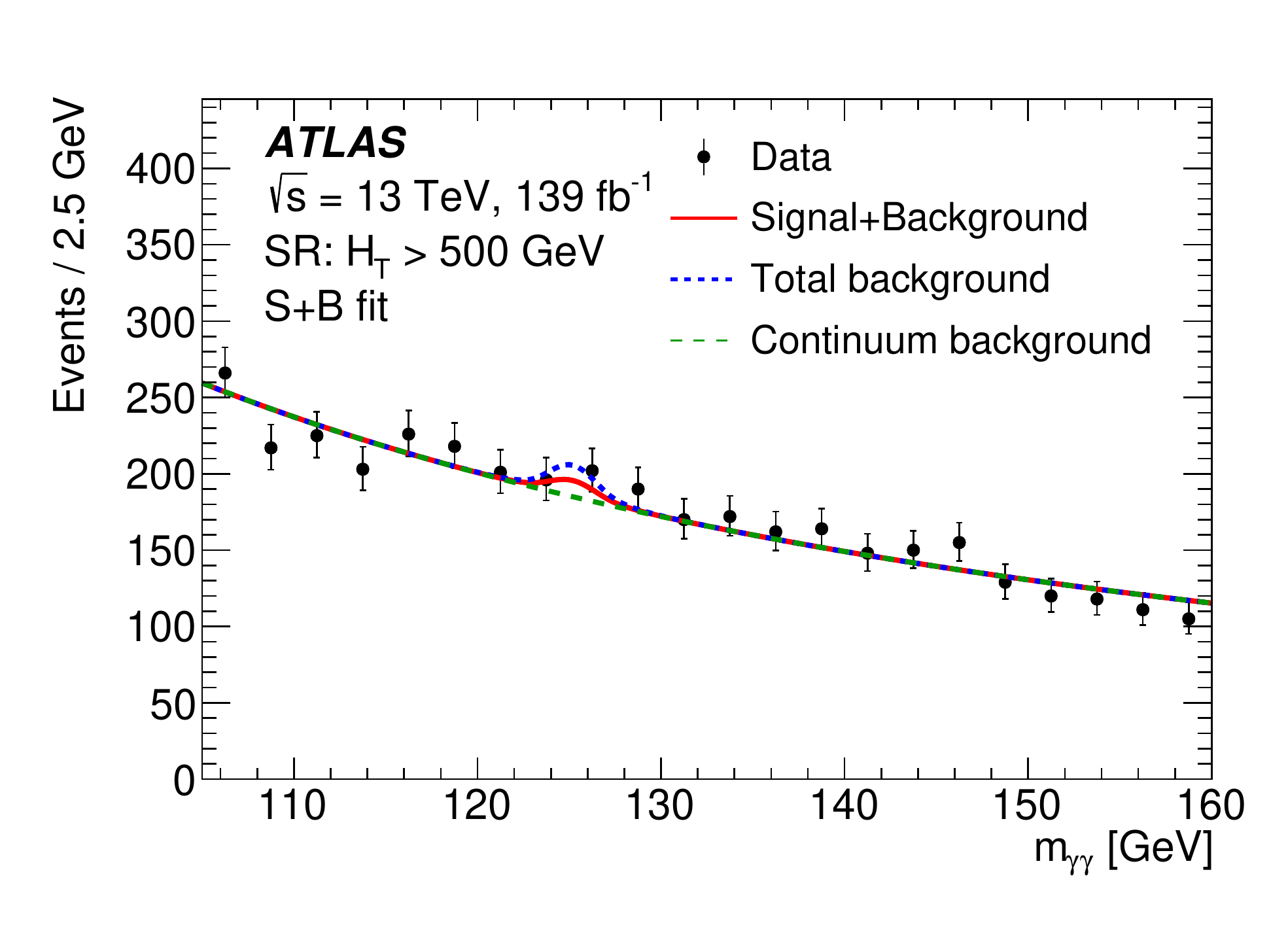}}
\subfloat[\HT$>$\,1000 \GeV]{\includegraphics[width=0.33\textwidth]{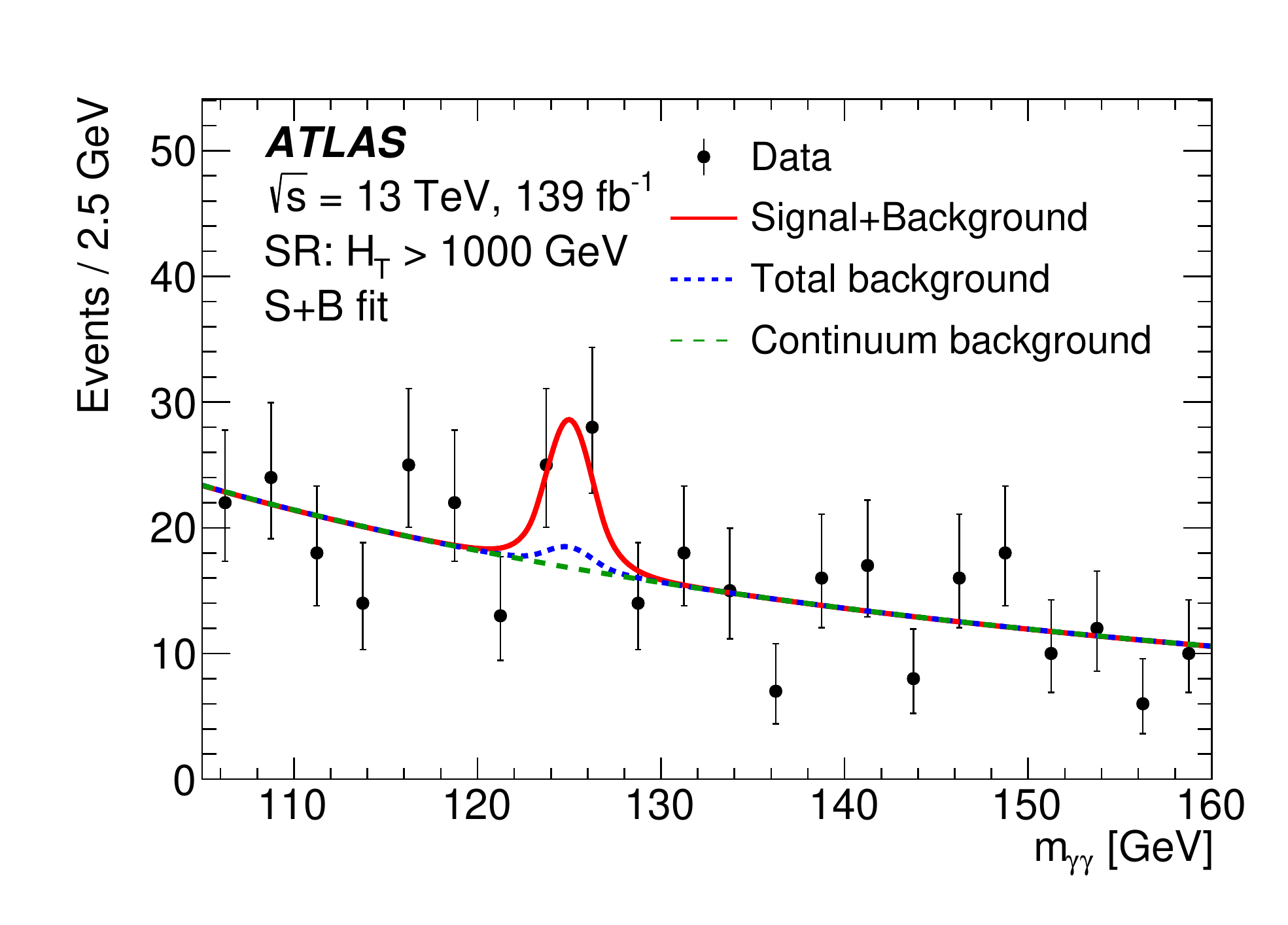}}
\subfloat[\HT$>$\,1500 \GeV]{\includegraphics[width=0.33\textwidth]{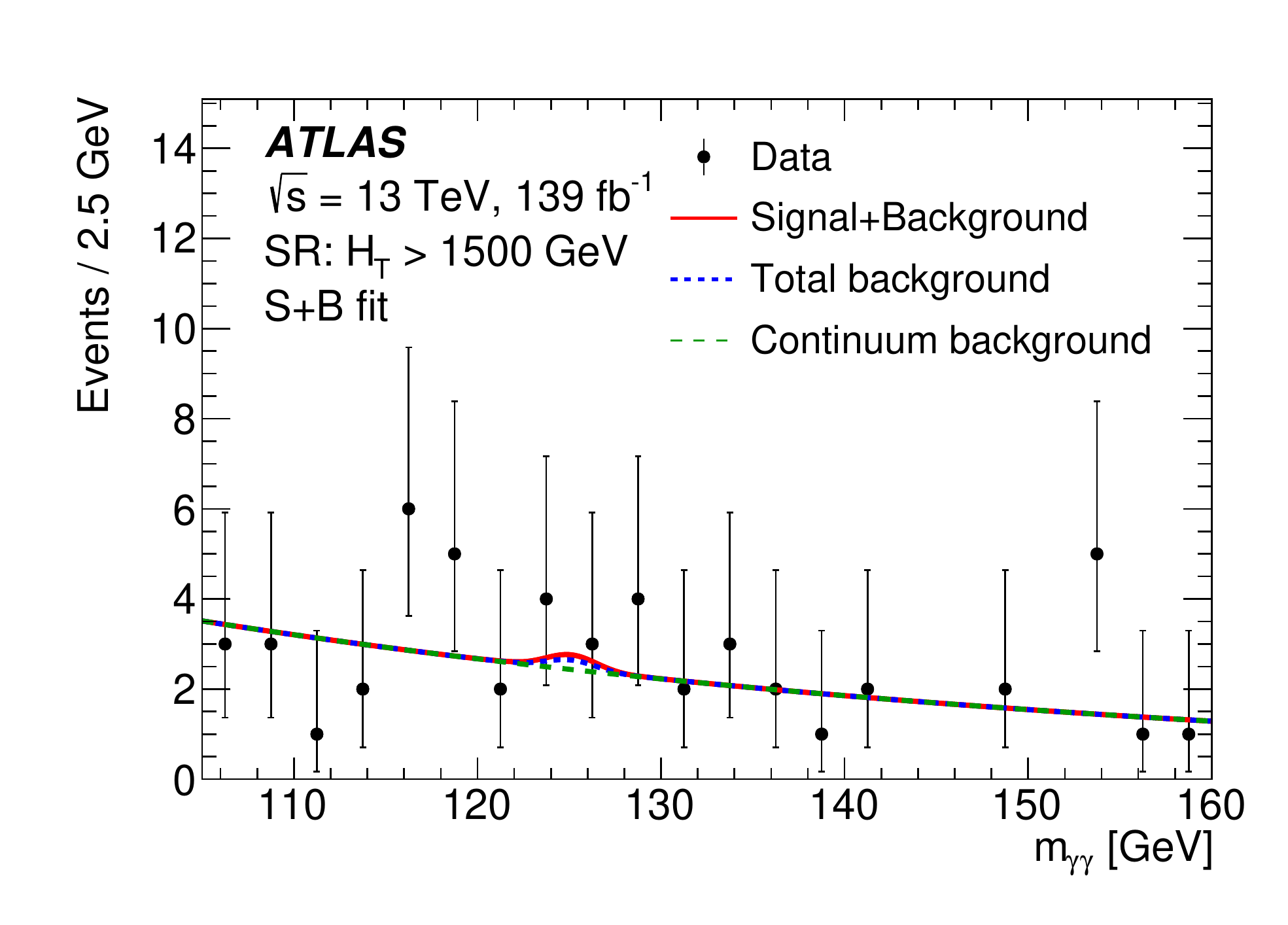}}
 
\caption{Diphoton mass distributions for high-jet-activity signal regions.
Data are shown together with the signal-plus-background fit. The overall fitted signal-plus-background pdf is shown as a solid red curve, the contributions from the resonant Higgs boson and continuum backgrounds are shown as a dashed blue curve, while the contributions only from the continuum backgrounds are shown as a dashed green curve. The difference between the solid red and the dashed blue curves indicates the fitted deviation from the SM expectation. }
\end{figure}
 
\begin{figure}[htbp]
\centering
\subfloat[\met$>$\,100 \GeV]{\includegraphics[width=0.33\textwidth]{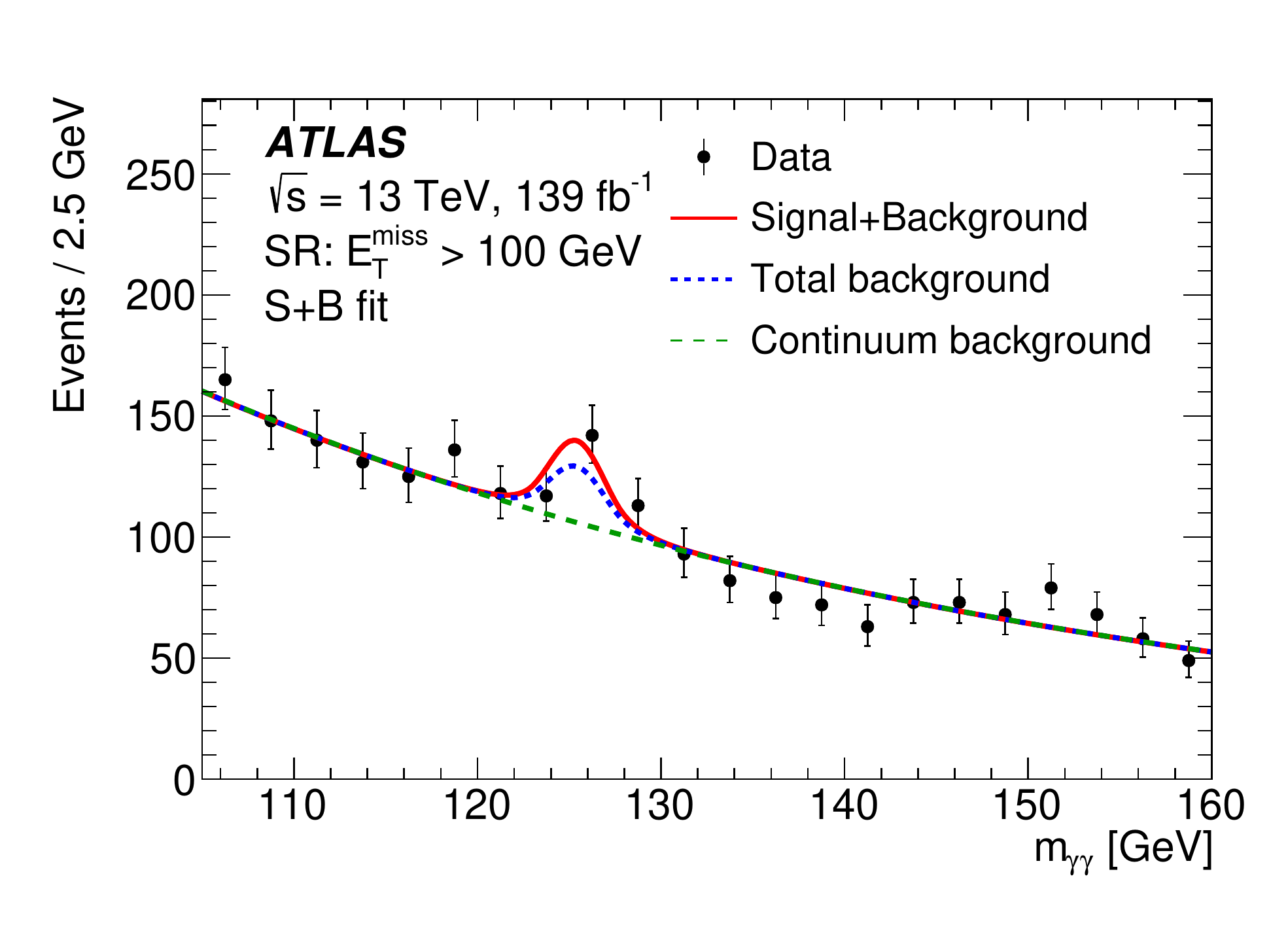}}
\subfloat[\met$>$\,200 \GeV]{\includegraphics[width=0.33\textwidth]{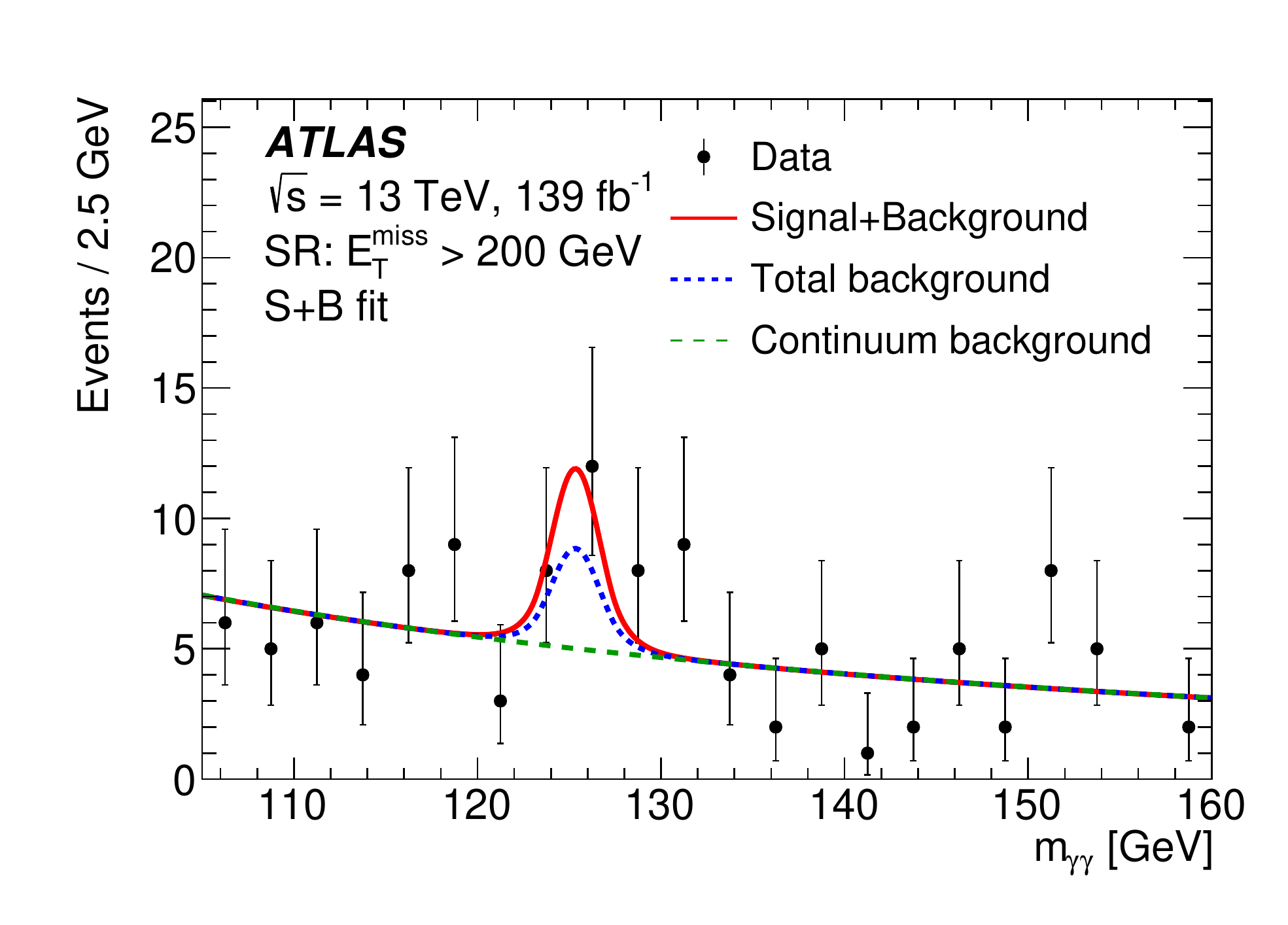}}
\subfloat[\met$>$\,300 \GeV]{\includegraphics[width=0.33\textwidth]{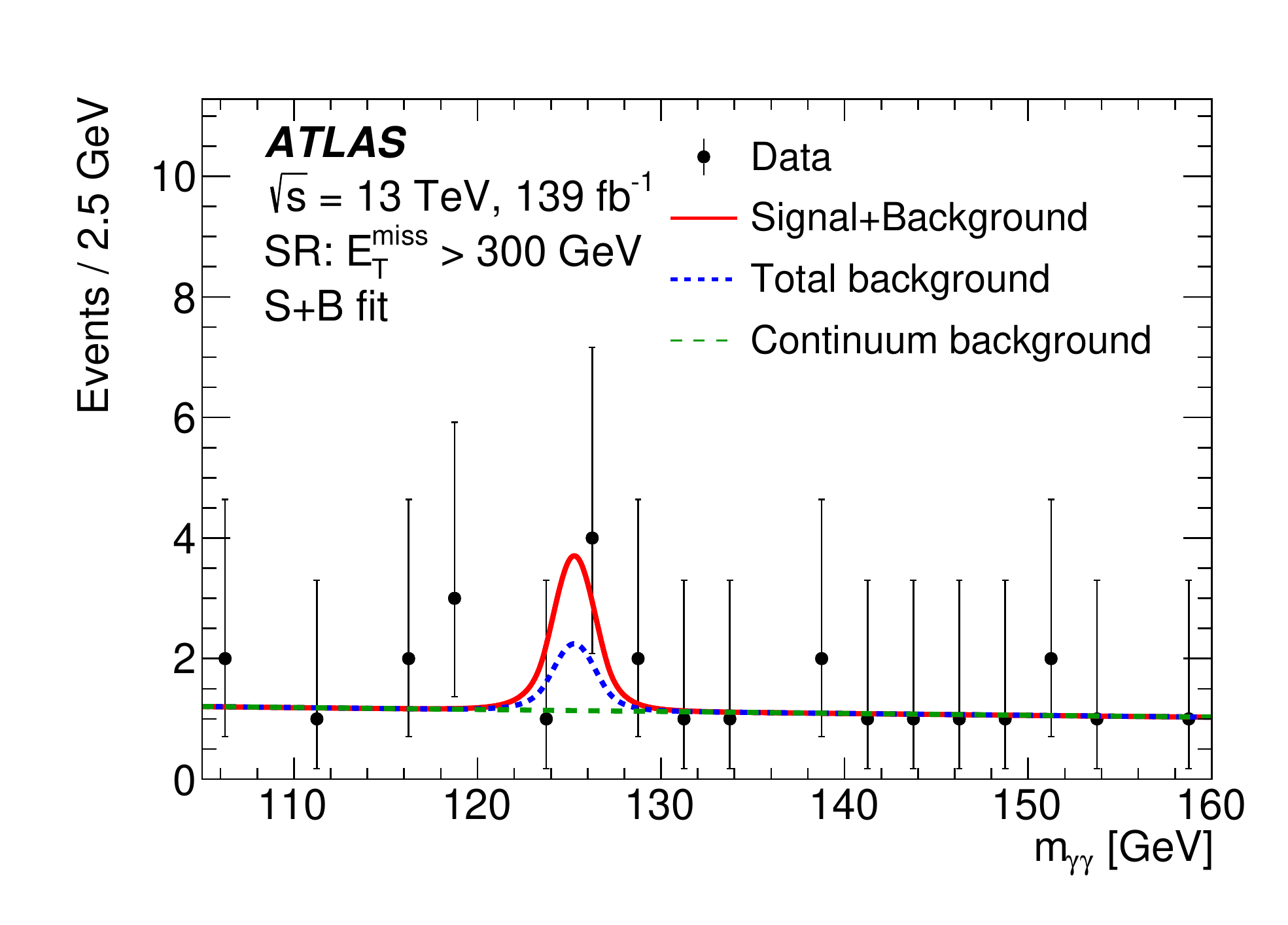}}\\
 
\caption{Diphoton mass distributions for \met\ signal regions.
Data are shown together with the signal-plus-background fit. The overall fitted signal-plus-background pdf is shown as a solid red curve, the contributions from the resonant Higgs boson and continuum backgrounds are shown as a dashed blue curve, while the contributions only from the continuum backgrounds are shown as a dashed green curve. The difference between the solid red and the dashed blue curves indicates the fitted deviation from the SM expectation. }
\end{figure}

\begin{figure}[htbp]
\centering
\subfloat[${\geq}1\ell$]{\includegraphics[width=0.33\textwidth]{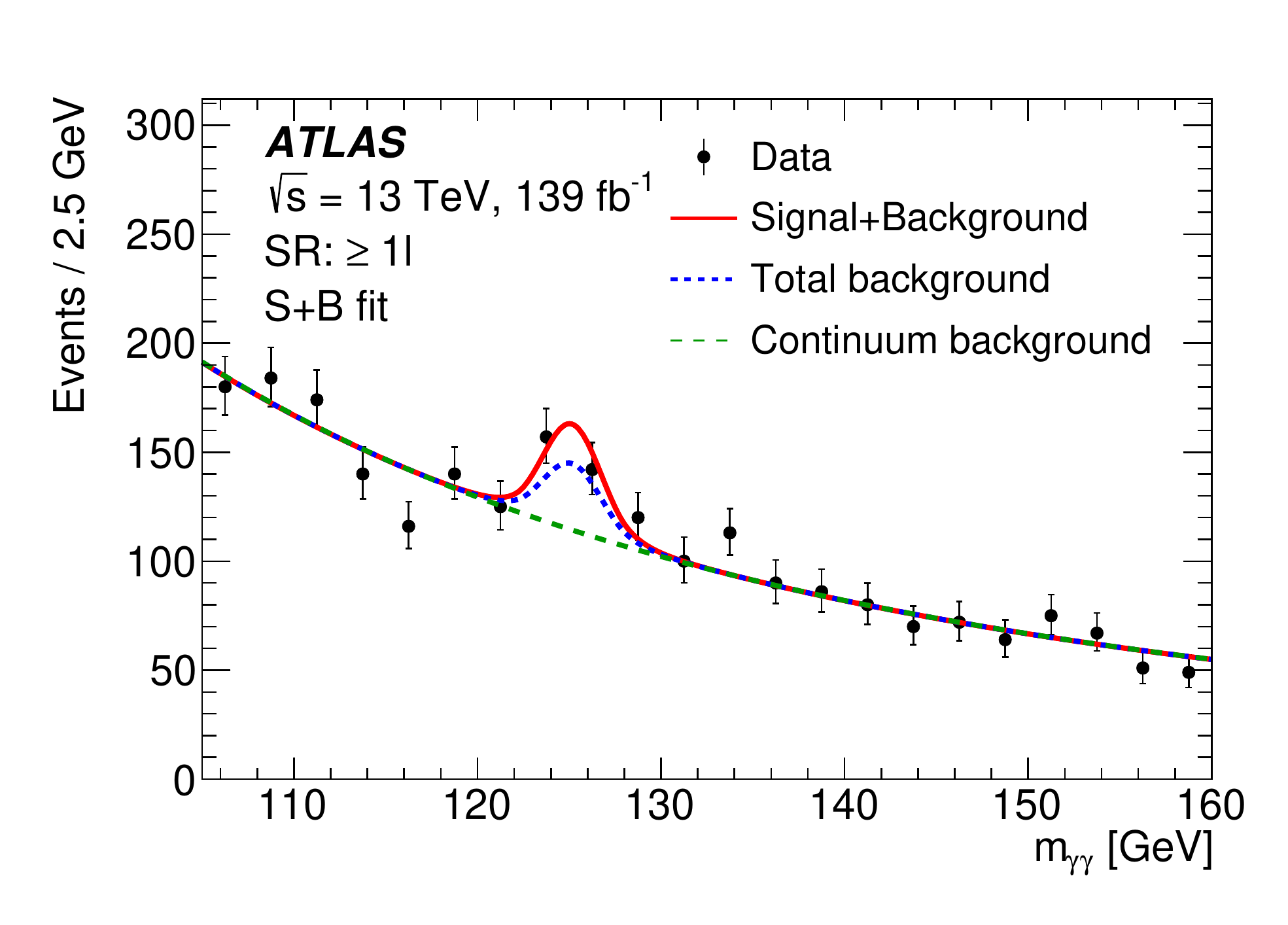}}
\subfloat[$2\ell$]{\includegraphics[width=0.33\textwidth]{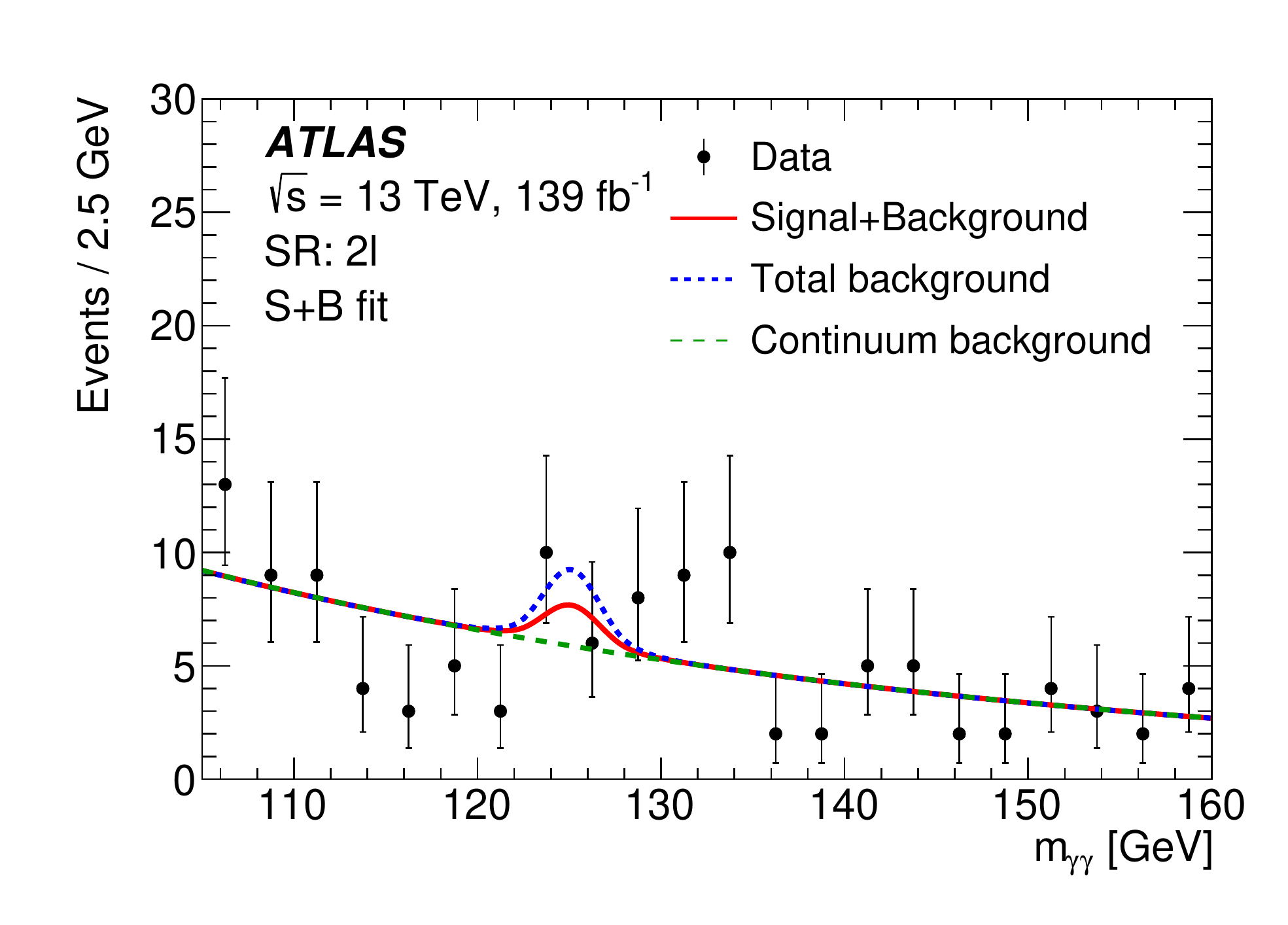}}
\subfloat[$2\ell$--$\cancel{Z}$]{\includegraphics[width=0.33\textwidth]{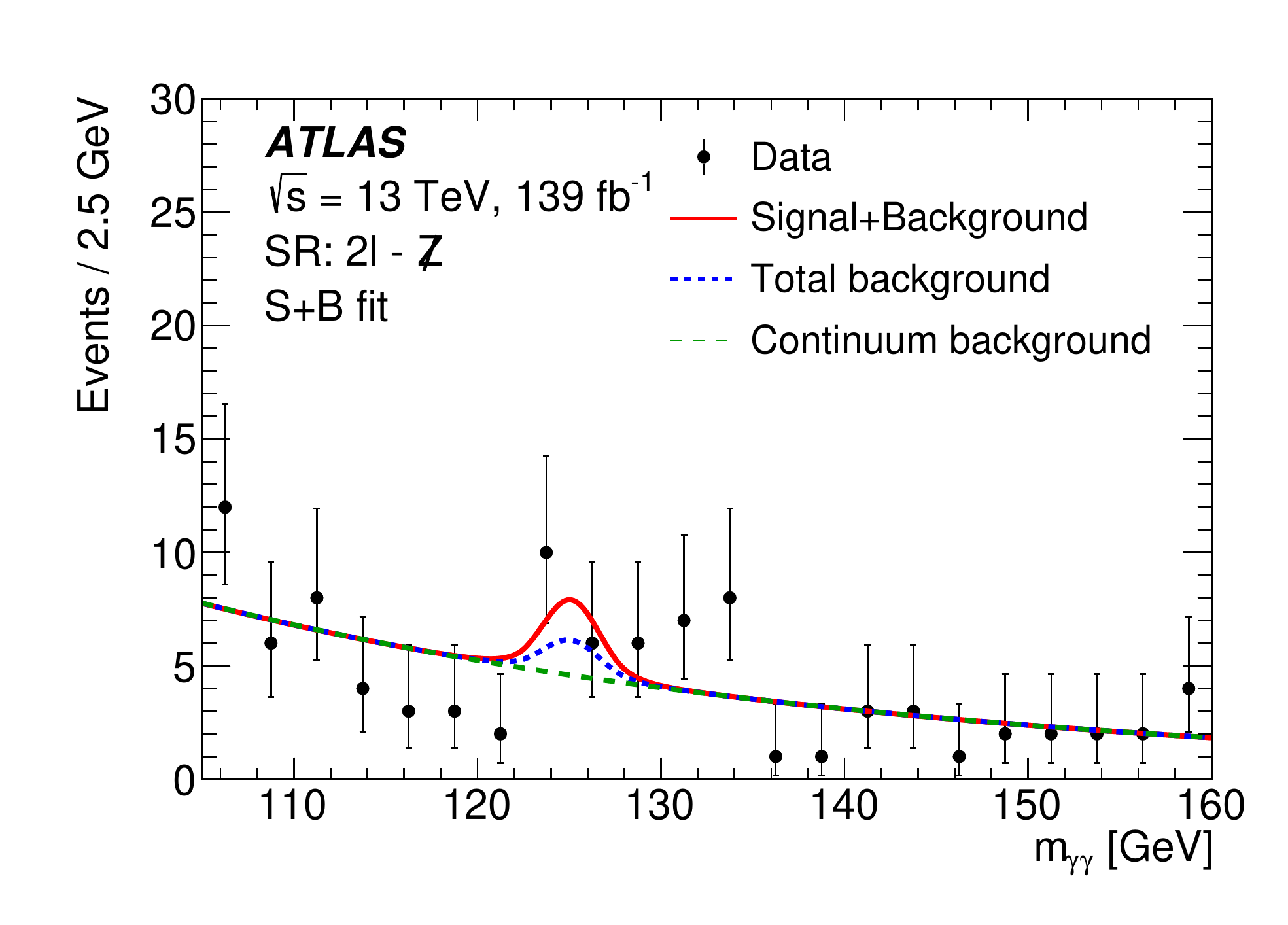}}\\
\subfloat[SS-$2\ell$]{\includegraphics[width=0.33\textwidth]{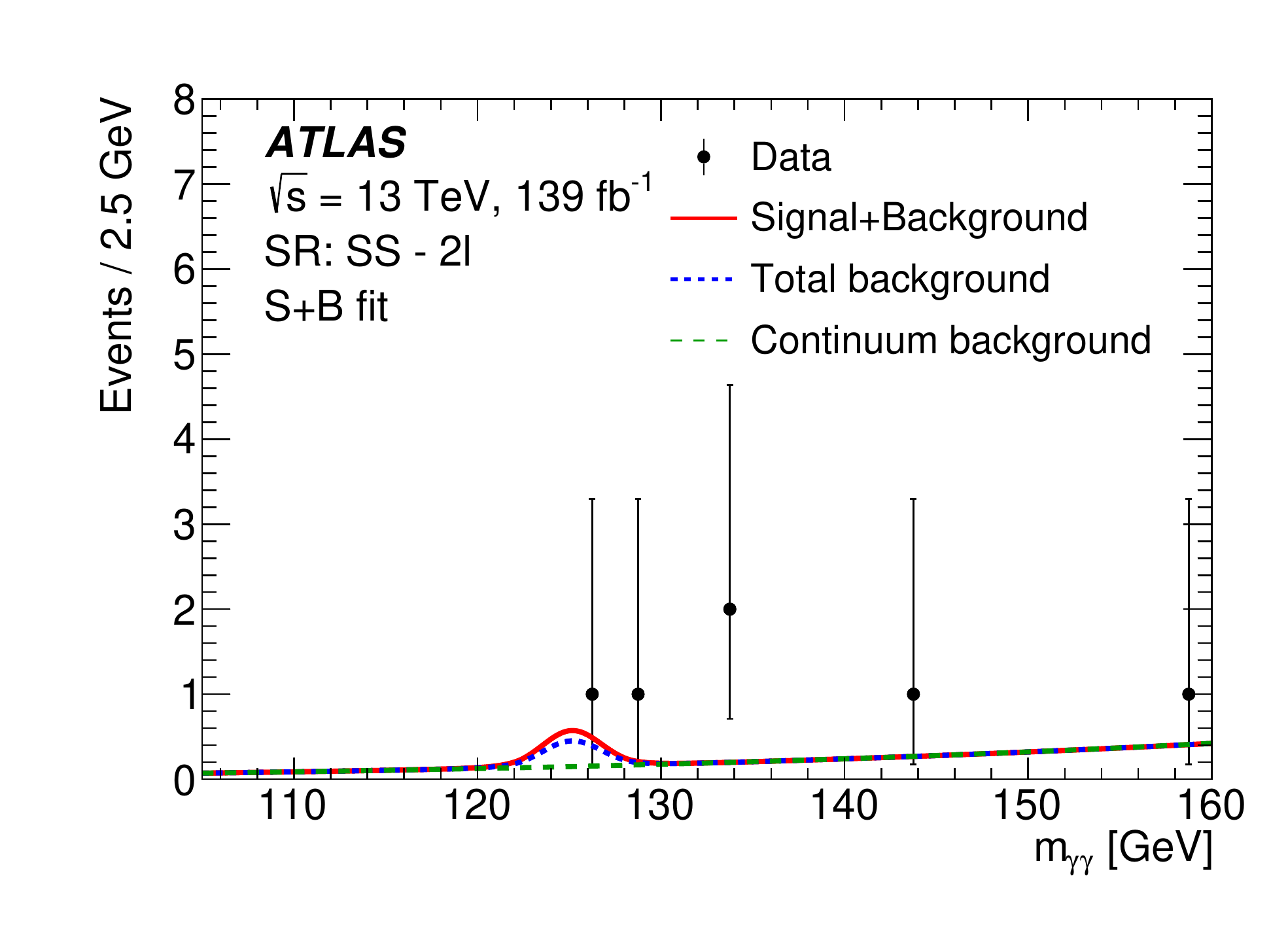}}
\subfloat[${\geq}2\tau$]{\includegraphics[width=0.33\textwidth]{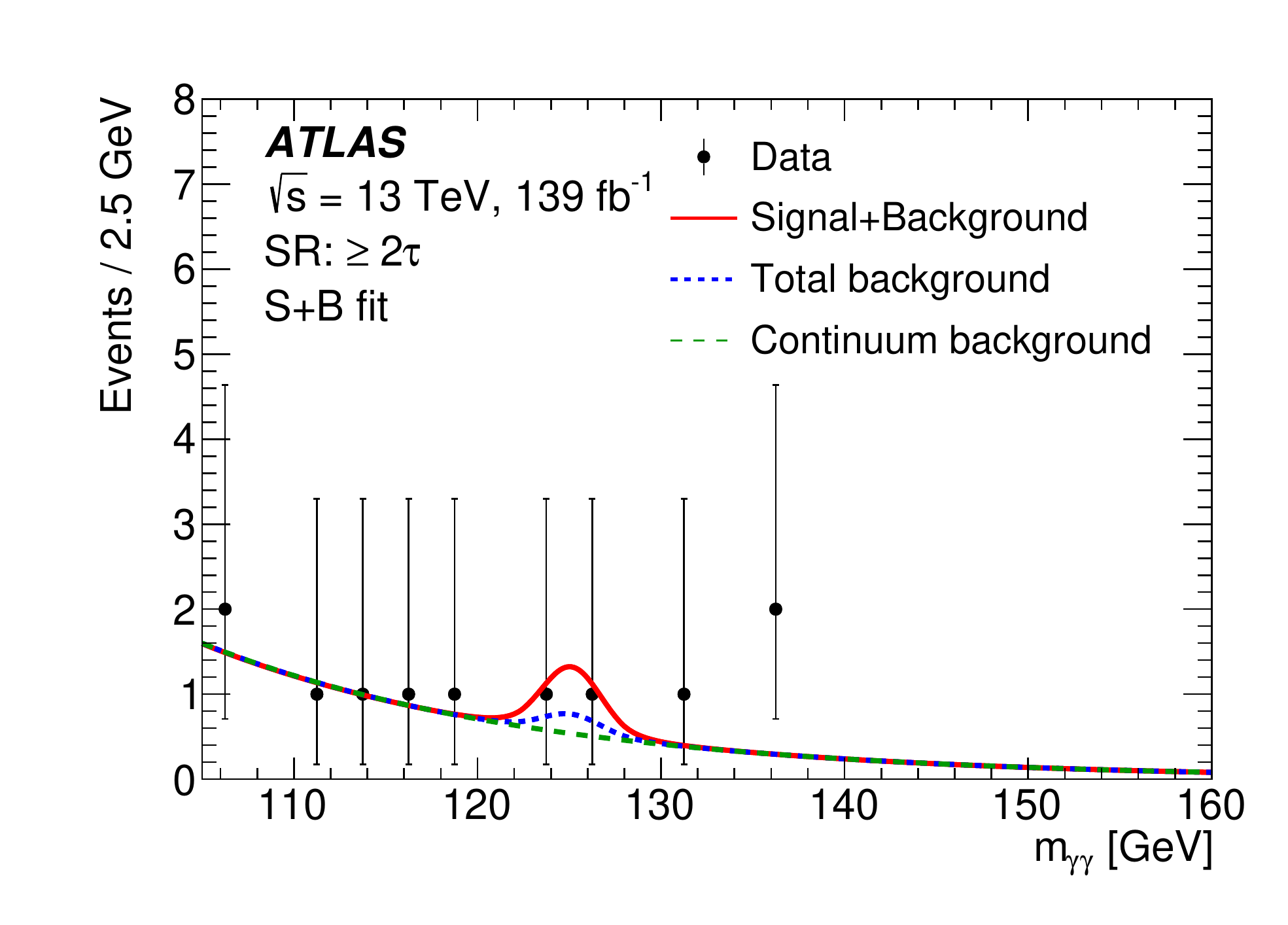}}
 
\caption{Diphoton mass distributions for lepton signal regions.
Data are shown together with the signal-plus-background fit. The overall fitted signal-plus-background pdf is shown as a solid red curve, the contributions from the resonant Higgs boson and continuum backgrounds are shown as a dashed blue curve, while the contributions only from the continuum backgrounds are shown as a dashed green curve. The difference between the solid red and the dashed blue curves indicates the fitted deviation from the SM expectation. }
\end{figure}

\begin{figure}[htbp]
\centering
\subfloat[$\ell b$]{\includegraphics[width=0.33\textwidth]{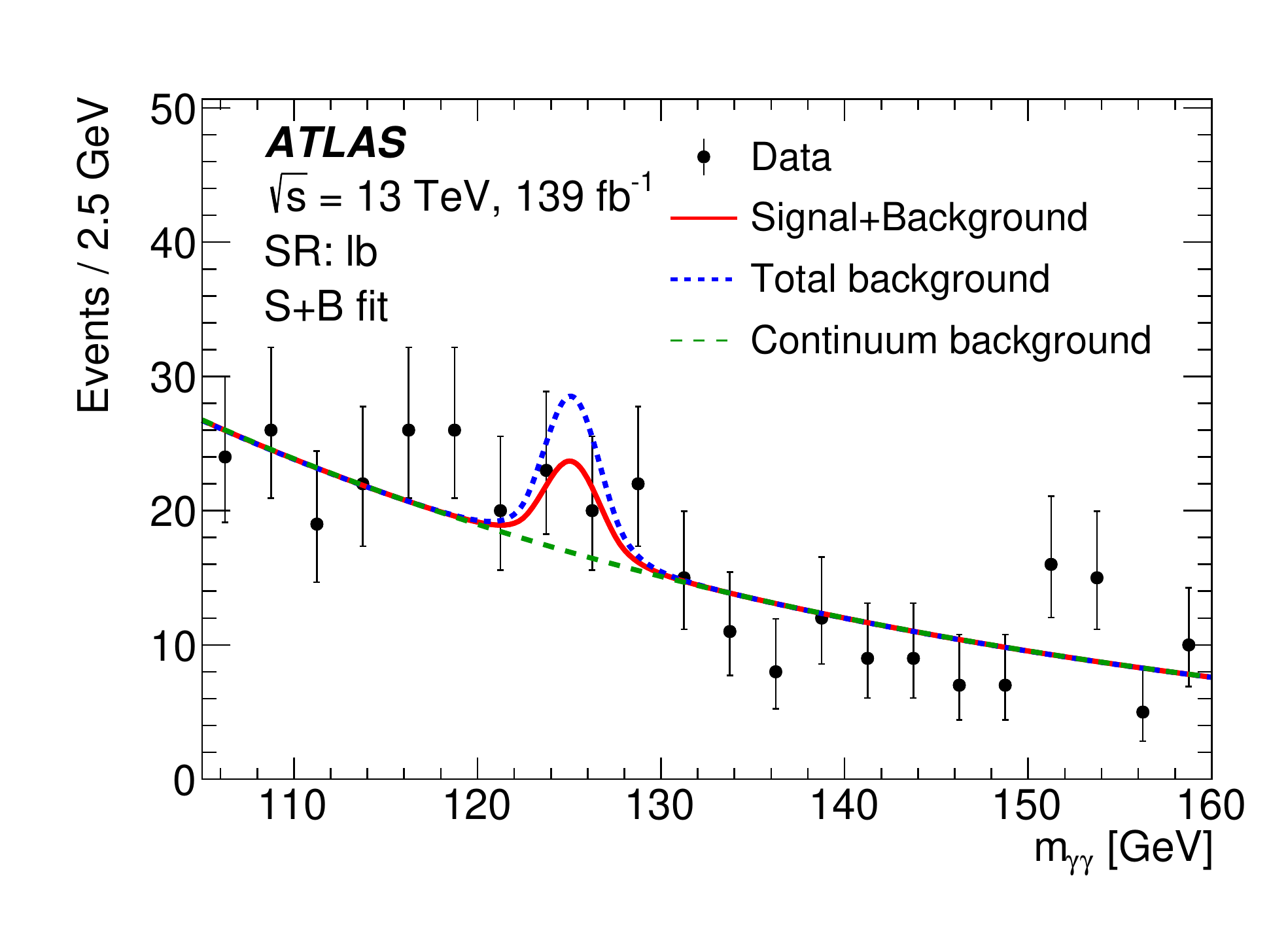}}
\subfloat[$t_\text{lep}$]{\includegraphics[width=0.33\textwidth]{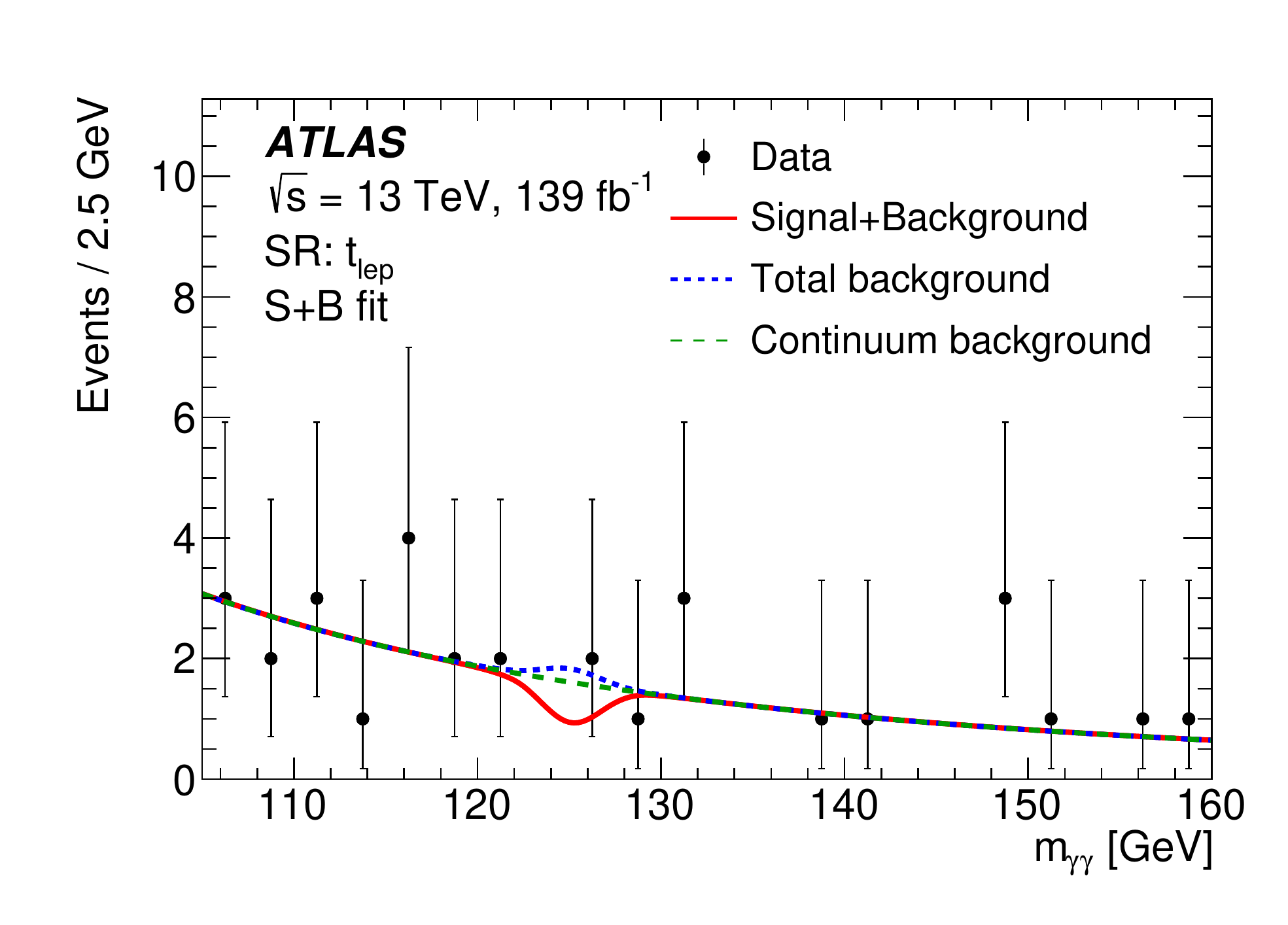}}
\subfloat[$t_\text{had}$]{\includegraphics[width=0.33\textwidth]{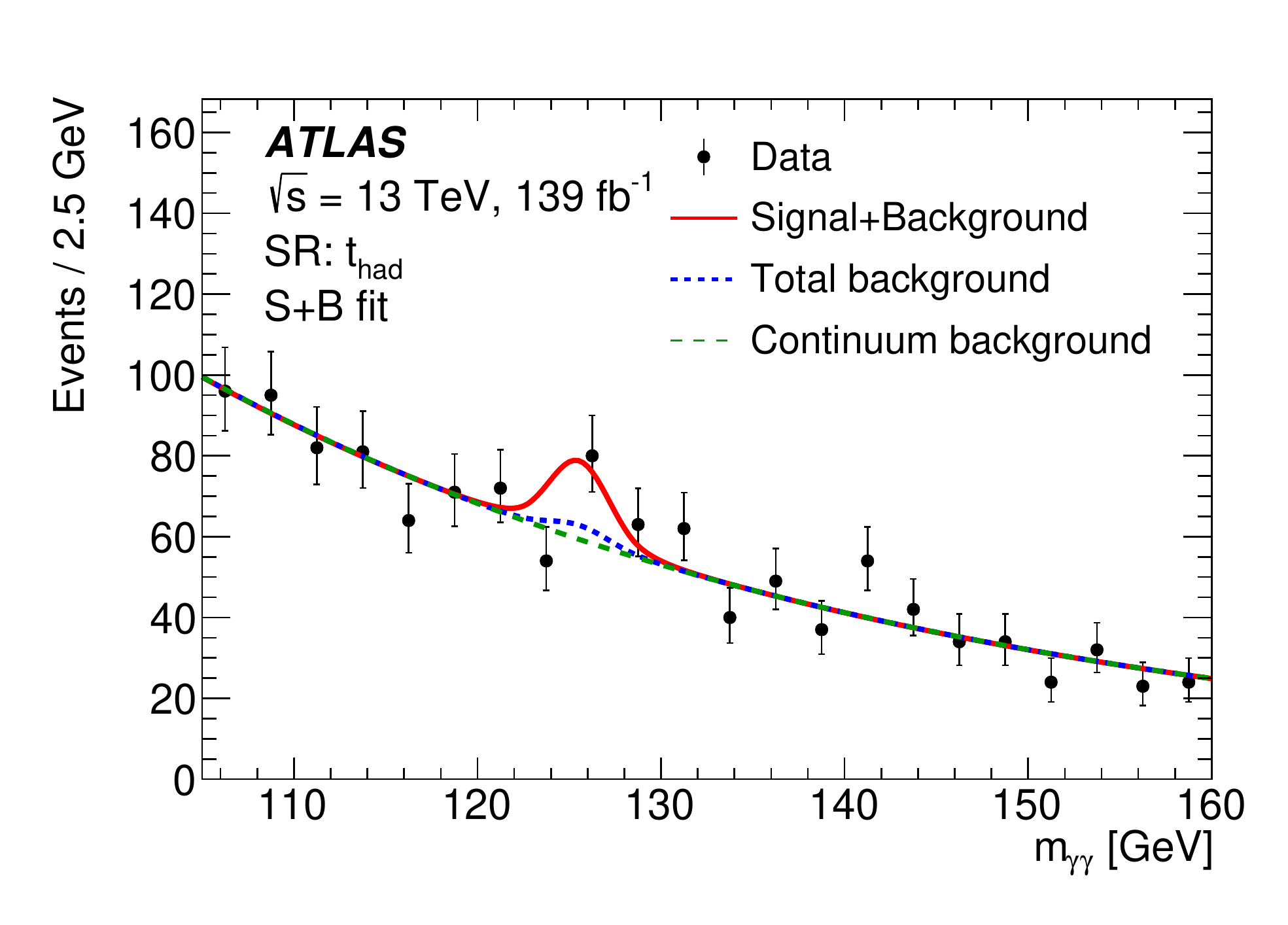}}\\
 
\caption{Diphoton mass distributions for top-quark signal regions.
Data are shown together with the signal-plus-background fit. The overall fitted signal-plus-background pdf is shown as a solid red curve, the contributions from the resonant Higgs boson and continuum backgrounds are shown as a dashed blue curve, while the contributions only from the continuum backgrounds are shown as a dashed green curve. The difference between the solid red and the dashed blue curves indicates the fitted deviation from the SM expectation. }
\end{figure}
 
\FloatBarrier

\subsection{Upper limits on BSM production cross sections}
\begin{figure}[htbp]
\centering
\includegraphics[width=0.95\textwidth]{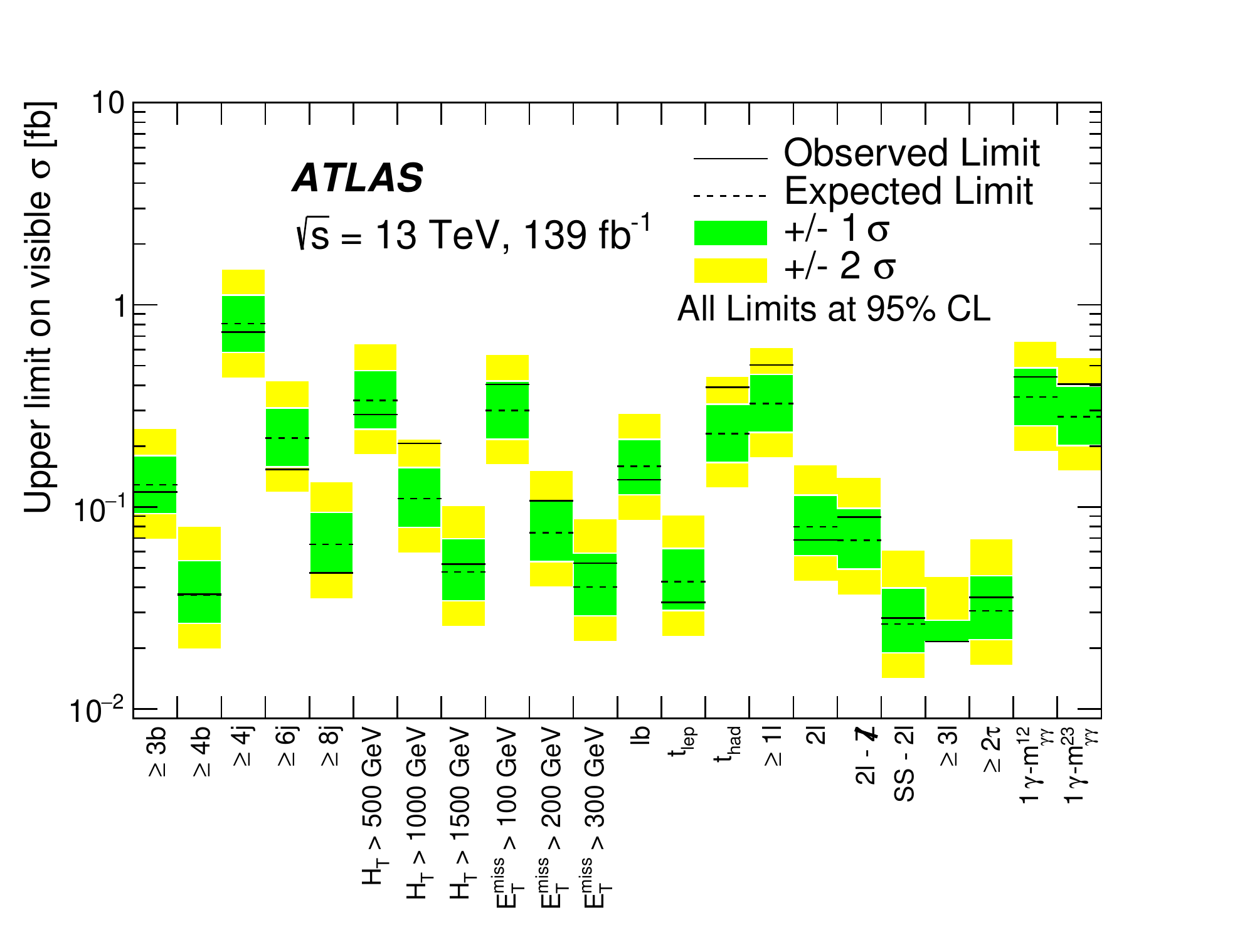}
\caption{Observed and expected 95\% CL limits on the visible cross section of BSM Higgs boson production in each signal region. The $1\sigma$ and $2\sigma$ variations of the expected limit are also shown as green and yellow bands, respectively.}
\label{fig:limit_xsec}
\end{figure}

Since no significant excess beyond the SM expectation was observed, the \mgg\ distributions in the signal regions, as well as the event count in the multilepton (${\geq}3\ell$) region, are used to set limits on the visible cross section of BSM production of the Higgs boson with a mass of 125.09 GeV in these regions.
The 95\% CL limits were derived using the modified frequentist CL$_\text{s}$ method~\cite{Read:2002hq}. The limits are shown in Table~\ref{tab:interp_expected_xsc} and graphically in  Figure~\ref{fig:limit_xsec}.
The observed 95\% CL limit on the visible cross section ranges from 0.05~fb (${\ge}3\ell$) to 0.7~fb ($\ge$4j).

The observed limits reported here are correlated since the signal regions overlap. To enable the use of limits from multiple signal regions to constrain BSM processes, the correlation between the observed limits from any pair of signal regions considered in this paper is estimated using a pseudo-experiment-based bootstrapping procedure~\cite{ATL-PHYS-PUB-2021-011}. Readers of this paper could derive a combined limit on the cross section of a BSM signal using the observed limits in individual signal regions and their correlations. Only regions with at least 50 events in the \mgg\ range from 105~\GeV\ to 160~\GeV\ were checked using this procedure. Correlations larger than 5\% are observed for a few regions and are reported in Table~\ref{tab:limit:correlation}.

\begin{table}[h!]
\centering
\caption{Observed and expected 95\% CL upper limits on the visible cross section of BSM Higgs boson production in each signal region. The $\pm 1\sigma$ and $\pm 2\sigma$ variations of the expected limit are also shown.}
\label{tab:interp_expected_xsc}
\scriptsize
\begin{tabular}{lcccccc}
\hline
& \multicolumn{6}{c}{95\% CL limit [fb]} \\
SR & Observed & Median & $+1\sigma$ & $-1\sigma$ & $+2\sigma$ & $-2\sigma$ \\
\hline
\hline
\multicolumn{7}{c}{Heavy flavour}\\
\hline
 
$\geq$3$b$ & 0.12\phantom{0} & 0.13\phantom{0} & 0.18\phantom{0} & 0.093 & 0.25\phantom{0} & 0.069 \\
$\geq$4$b$ & 0.037 & 0.037 & 0.054 & 0.026 & 0.081 & 0.02\phantom{0} \\
 
\hline
\multicolumn{7}{c}{High jet activity}\\\hline
$\geq$4j & 0.73\phantom{0} & 0.81\phantom{0} & 1.1\phantom{00} & 0.58\phantom{0} & 1.5\phantom{00} & 0.43\phantom{0} \\
$\geq$6j & 0.15\phantom{0} & 0.22\phantom{0} & 0.31\phantom{0} & 0.16\phantom{0} & 0.42\phantom{0} & 0.12\phantom{0} \\
$\geq$8j & 0.047 & 0.065 & 0.094 & 0.047 & 0.13\phantom{0} & 0.035 \\
\HT$>$\,500 \GeV & 0.28\phantom{0} & 0.34\phantom{0} & 0.47\phantom{0} & 0.24\phantom{0} & 0.64\phantom{0} & 0.18\phantom{0} \\
\HT$>$\,1000 \GeV & 0.21\phantom{0} & 0.11\phantom{0} & 0.16\phantom{0} & 0.079 & 0.22\phantom{0} & 0.059 \\
\HT$>$\,1500 \GeV & 0.052 & 0.048 & 0.07\phantom{0} & 0.034 & 0.1\phantom{00} & 0.026 \\
 
\hline
\multicolumn{7}{c}{\met}\\\hline
\met$>$\,100 \GeV & 0.4\phantom{00} & 0.3\phantom{00} & 0.42\phantom{0} & 0.22\phantom{0} & 0.57\phantom{0} & 0.16\phantom{0} \\
\met$>$\,200 \GeV & 0.11\phantom{0} & 0.075 & 0.11\phantom{0} & 0.054 & 0.15\phantom{0} & 0.04\phantom{0} \\
\met$>$\,300 \GeV & 0.053 & 0.04\phantom{0} & 0.059 & 0.029 & 0.088 & 0.022 \\
\hline
\multicolumn{7}{c}{Top}\\ \hline
$\ell b$ & 0.14\phantom{0} & 0.16\phantom{0} & 0.22\phantom{0} & 0.11\phantom{0} & 0.29\phantom{0} & 0.085 \\
$t_\text{lep}$ & 0.034 & 0.043 & 0.062 & 0.031 & 0.092 & 0.023 \\
$t_\text{had}$ & 0.39\phantom{0} & 0.23\phantom{0} & 0.32\phantom{0} & 0.17\phantom{0} & 0.44\phantom{0} & 0.12\phantom{0} \\
\hline
\multicolumn{7}{c}{Lepton}\\\hline
$\geq$1$\ell$ & 0.5\phantom{00} & 0.32\phantom{0} & 0.45\phantom{0} & 0.23\phantom{0} & 0.61\phantom{0} & 0.17\phantom{0} \\
2$\ell$ & 0.069 & 0.08\phantom{0} & 0.11\phantom{0} & 0.057 & 0.16\phantom{0} & 0.043 \\
2$\ell$--$\cancel{Z}$ & 0.089 & 0.068 & 0.098 & 0.049 & 0.14\phantom{0} & 0.037 \\
SS-2$\ell$ & 0.028 & 0.026 & 0.04\phantom{0} & 0.019 & 0.061 & 0.014 \\
$\geq$3$\ell$ & 0.022 & 0.022 & 0.028 & 0.022 & 0.046 & 0.022 \\
$\geq$2$\tau$ & 0.036 & 0.031 & 0.046 & 0.022 & 0.07\phantom{0} & 0.016 \\
 
\hline
\multicolumn{7}{c}{Photon}\\\hline
1$\gamma$--$m_{\gamma\gamma}^{12}$ & 0.44\phantom{0} & 0.35\phantom{0} & 0.49\phantom{0} & 0.25\phantom{0} & 0.66\phantom{0} & 0.19\phantom{0} \\
1$\gamma$--$m_{\gamma\gamma}^{23}$ & 0.4\phantom{00} & 0.28\phantom{0} & 0.4\phantom{00} & 0.2\phantom{00} & 0.55\phantom{0} & 0.15\phantom{0} \\
 
\hline
\end{tabular}
\end{table}
 
\FloatBarrier
 
\begin{table}[h!]
\centering
\caption{Correlations between observed 95\% CL limits for pairs of signal regions. Only regions with a correlation greater than 5\% are reported. Pairs of regions that appear in previous rows are not repeated.}
\label{tab:limit:correlation}
\begin{tabular}{c|c}
\hline
Region & Region (Correlation) \\
\hline
$\geq$4j  & $\geq$6j (29\%), \HT$>$\,500 \GeV\ (28\%)  \\
$\geq$6j  & $\geq$8j (25\%), \HT$>$\,500 \GeV\ (26\%), \HT$>$\,1000 \GeV\ (8\%)\\
\HT$>$\,500 \GeV & \HT$>$\,1000 \GeV\ (32\%) \\
\met$>$\,200 GeV & \met$>$\,100 \GeV\ (29\%) \\
$1\ell$ & $2\ell$ (26\%), $\ell{b}$ (36\%) \\
$2\ell$ & $2\ell$--$\cancel{Z}$ (97\%) \\
\hline
\end{tabular}
\end{table}
 
\subsection{Interpretation with particle-level simulation}
\label{sec:result:interpretation}
This paper aims to present results that can be used without detailed knowledge of the ATLAS detector to derive constraints on the BSM processes of interest. For a specific BSM process, its particle-level cross section, \sigmatruth, in a given region can be derived from MC simulation with particle-level requirements as given in Table~\ref{tab:regsummary}. To determine if the BSM process is excluded or not, \sigmatruth\ should be compared with a particle-level 95\% CL limit, \sigmapart, for the same region. Since this paper only reports a 95\% CL limit on the visible cross section at the detector level, \sigmadet, for any of the signal regions, a signal-region-specific detector efficiency, $\epsilon$, must be provided to convert \sigmadet\ to \sigmapart. More specifically, one has $\sigmapart=\sigmadet/\epsilon$.

The region-specific detector efficiency $\epsilon$ is defined as the ratio of the event yield in a signal region at the detector level to that in a signal region defined at the particle level. As such, for each signal region considered in this paper, a particle-level definition must be specified. Table~\ref{tab:truthdefs} summarizes the particle-level object definitions. Particle-level jets are built from all MC generator-level \enquote{truth} particles, removing the contributions from muons and stable particles such as neutrinos and lightest supersymmetric particles, using the \antikt algorithm with a radius parameter of 0.4. The particle-level objects are required to be isolated by implementing the following overlap removal procedure: electrons, muons, photons, and $\tau$-leptons are removed if there is another object of the same type within $\Delta R < 0.1$. Then any muon, jet or electron is removed if it is  within $\Delta R = 0.4$ of any photon, any jet  within $\Delta R = 0.2$ of any electron is removed, and finally any muon or electron within $\Delta R = 0.4$ of any jet is removed. In this paper, a particle-level $b$-jet is defined as a jet that is within $\Delta R = 0.4$ of a $b$-quark, and the $b$-quark should pass the requirements of $\pt > 25$~\GeV\ and $|\eta| < 2.5$. This requirement is not fully efficient due to the kinematic requirements on the $b$-quark. Depending on the specific signal process, the requirement has an efficiency of matching a particle-level jet to a $b$-quark at the level of 70\% to 80\%, for a $b$-quark with $\pt > 25$~\GeV\ and $|\eta| < 2.5$.  Then, the particle-level signal regions can be defined with the particle-level objects, and the definitions are detailed in Table~\ref{tab:regsummary}.
 
\begin{table}[th!]
\begin{center}\renewcommand\arraystretch{1.2}
\caption{Particle-level object definitions and event preselection.}
\label{tab:truthdefs}
\begin{tabular}{l|c}
\hline
Object & Requirements\\
\hline
\hline
Photon & $\pt > 22~\GeV, |\eta| < 2.5$ \\
Electron &  $\pt > 10$~\GeV , $|\eta | < 2.5$  \\
Muon & $\pt > 10$~\GeV , $|\eta | < 2.7$ \\
Jet & $\pt > 25$~\GeV , $|\eta | < 4.4$  \\
$b$-jet & $\pt > 25$~\GeV , $|\eta | < 2.5$, $\Delta R(b\textrm{-quark, jet}) < 0.4$ \\
$\tau$-lepton & $\pt > 20$~\GeV , $|\eta | < 2.5$ \\
\hline
\multicolumn{2}{c}{Preselection}\\
\hline
\hline
\multicolumn{2}{c}{$\pt^{\gamma1} > 35$~\GeV, $\pt^{\gamma2} > 25$~\GeV, $\pt^{\gamma1} / m_{\gamma \gamma}>0.35$, $\pt^{\gamma2} / m_{\gamma \gamma}>0.25$}\\
\hline
\end{tabular}
\end{center}
\end{table}
\FloatBarrier
 
The detector efficiency, $\epsilon$, is evaluated for all signal regions considered in this paper and is summarized in Table~\ref{tab:updated_epsilon_range}. For each signal region, the value of $\epsilon$ is calculated for a few SM and BSM Higgs boson production processes, and the range of $\epsilon$ values is reported. Since the final-state object multiplicity and kinematics can differ between processes, variations in $\epsilon$ are expected. The reported $\epsilon$ values are subject to a small statistical uncertainty, which is much smaller than the variation in $\epsilon$ between different physics processes. In general, the detector-level selection is always less efficient than the particle-level selection, except for the following two situations. First, when selecting a particle-level $b$-jet, the $\Delta{R}$ matching procedure may be less efficient than the detector-level $b$-tagging, for which a $b$-jet is defined as a jet containing a $b$-hadron. The adoption of the $\Delta{R}$ matching procedure is motivated by the convenience of implementation at the particle level. Second, when a signal region is defined by placing a cut at a value in the tail of a variable's distribution, such as for \HT\ or the number of jets, detector-level effects could lead to event migration and hence a higher detector-level selection efficiency. However, when all the detector effects are combined, as shown in Table~\ref{tab:updated_epsilon_range}, $\epsilon$ is always less than 1. The value of $\epsilon$ is very small for the $2\tau$ signal region because the particle-level $\tau$-leptons include all $\tau$-leptons regardless of their decay modes. This choice is also motivated by the convenience of selection implementation at the particle level.

As an example, an electroweak SUSY production model, corresponding to the one shown in the first row of Table~\ref{tab:samples:bsm} is used to demonstrate how information reported in this section can be used to derive a constraint on a given BSM model. In this SUSY process, a chargino and neutralino pair ($\tilde{\chi}_1^\pm \tilde{\chi}_2^0$) is produced in the $pp$ collision, which results in a final state of a $W$ boson, a Higgs boson, and two light neutralinos, $\tilde{\chi}_1^0$. The light neutralinos are undetectable, giving rise to large values of \met. The masses of the pair-produced chargino and neutralino are 150~\GeV, and the light neutralino's mass is 0.5~\GeV. For this model, a targeted search has already been carried out by the ATLAS Collaboration~\cite{SUSY-2018-23}, which reported a 95\% CL upper limit of 2~pb for the cross section at this specific signal point.
 
The $\met > 100$~\GeV\ signal region can be used to probe the same SUSY process, which predicts a particle-level cross section, \sigmatruth, of 1.05~fb for this $\met > 100$~\GeV\ region.
The observed 95\% CL upper limit on the BSM cross section at the detector level is $\sigmadet = 0.4$~fb, for the $\met > 100$~\GeV\ region, and its corresponding $\epsilon$ is chosen, conservatively, to be 0.6, according to Table~\ref{tab:updated_epsilon_range}. The particle-level 95\% CL limit on the BSM cross section in the $\met > 100$~\GeV\ region is therefore $\sigmapart = 0.67$~fb, which would exclude the SUSY model considered here. The observed limit in the $\met >100$~\GeV\ region in the \Hgg\ final state can be converted to a limit on the inclusive cross section for this SUSY model by taking into account the selection acceptance at the particle level and the \Hgg\ branching ratio, and the resulting inclusive cross-section limit is 3.3~pb. This limit is of the same order of magnitude as the 2~pb upper limit in the dedicated search, but is 65\% weaker.

\begin{table}[bh!]
\begin{center}
\caption{Range of \eps\ values for all SRs. The detector efficiency \eps\ is only calculated for processes where the reconstruction-level yield in a given signal region is greater than 0.1~events for 139~\ifb\ of data. }
\label{tab:updated_epsilon_range}
\resizebox{\columnwidth}{!}{
\begin{tabular}{lcc}
\hline
 
SR & Relevant processes & Range of $\epsilon$ \\ \hline
\hline
\multicolumn{3}{c}{Heavy flavour}\\
\hline
 
$\geq$3$b$& \sbottom, \stopTwo& 0.68--0.81\\
$\geq$4$b$& \sbottom, \stopTwo& 0.64--0.97\\ 
\hline
\multicolumn{3}{c}{High jet activity}\\\hline
 
$\geq$4j& $t\bar{t}H$, \sbottom, \stopTwo, $tWH$ & 0.60--0.70\\
$\geq$6j& $t\bar{t}H$, \sbottom, \stopTwo, $tWH$ &  0.64--0.80\\
$\geq$8j& $t\bar{t}H$, \sbottom, \stopTwo & 0.65--0.90\\
\HT$>$\,500 \GeV& $t\bar{t}H$, \sbottom, \stopTwo, $tWH$ & 0.52--0.66\\
\HT$>$\,1000 \GeV& $t\bar{t}H$, \sbottom, \stopTwo, $tWH$ & 0.51--0.72\\
\HT$>$\,1500 \GeV& $t\bar{t}H$, \sbottom, \stopTwo & 0.41--0.73\\
\hline
\multicolumn{3}{c}{\met}\\\hline
\met$>$\,100 \GeV& $t\bar{t}H$, $tWH$, $WH$, $ZH$, \ConeNtwo &0.60--0.78\\
\met$>$\,200 \GeV& $t\bar{t}H$, $tWH$, $WH$, $ZH$, \ConeNtwo & 0.60--0.79\\
\met$>$\,300 \GeV&  $t\bar{t}H$, $tWH$, $WH$, $ZH$, \ConeNtwo &0.66--0.84\\
\hline
\multicolumn{3}{c}{Lepton}\\\hline
$\geq$1$\ell$& $WH$, $t\bar{t}H$, $tWH$, FCNC, \ConeNtwo & 0.40--0.48\\
2$\ell$& $ZH$, $t\bar{t}H$, \stopTwo, \ConeCone, \ConeNone & 0.21--0.48\\
2$\ell$--$\cancel{Z}$& $t\bar{t}H$, \stopTwo, \ConeCone, \ConeNone  &  0.20--0.46\\
${\geq}3\ell$ & \stopTwo, \ConeCone, \ConeNone & 0.18--0.33\\
SS-2$\ell$& \stopTwo, \ConeNone & 0.29--0.49\\
$\geq$2$\tau$& $ZH$, \stopTwo, \sbottom, \ConeCone, \ConeNone & 0.04--0.09\\
\hline
\multicolumn{3}{c}{Top quark}\\ \hline
$t_\text{lep}$& FCNC with semileptonically decaying top  & 0.32--0.36\\
$t_\text{had}$& FCNC with hadronically decaying top & 0.29--0.30\\
$\ell b$& $t\bar{t}H$, $tHjb$, $tWH$, FCNC with semileptonically decaying top & 0.41--0.52\\
\hline
\multicolumn{3}{c}{Photon}\\\hline
1$\gamma$--$m_{\gamma\gamma}^{12}$& \stopTwo, \sbottom &0.23--0.33\\
1$\gamma$--$m_{\gamma\gamma}^{23}$& \stopTwo& 0.35--0.40\\
\hline
\end{tabular}
}
\end{center}
\end{table}
\FloatBarrier

% End of text imported from the .//sections/results.tex input file

\section{Conclusions}
\label{sec:conclusion}

% The next lines are included from the .//sections/conclusions.tex input file
This paper reports a model-independent examination of the diphoton final states where a Higgs boson is produced, using a 139~\ifb\ data sample of 13~\TeV\ $pp$ collisions collected by the ATLAS detector during \RunTwo of the LHC. A total of 22 signal regions defined with various requirements on the additional reconstructed objects were examined, and no significant excess over the SM expectation was observed. Upper limits at 95\% CL were set on the visible cross section of BSM Higgs boson production in each of these signal regions. To enable interpretation of the results as a constraint on other models, a set of detector efficiency factors is also reported.

% End of text imported from the .//sections/conclusions.tex input file

\section*{Acknowledgements}

% The next lines are included from the .//acknowledgements/Acknowledgements.tex input file

We thank CERN for the very successful operation of the LHC, as well as the
support staff from our institutions without whom ATLAS could not be
operated efficiently.
 
We acknowledge the support of
ANPCyT, Argentina;
YerPhI, Armenia;
ARC, Australia;
BMWFW and FWF, Austria;
ANAS, Azerbaijan;
CNPq and FAPESP, Brazil;
NSERC, NRC and CFI, Canada;
CERN;
ANID, Chile;
CAS, MOST and NSFC, China;
Minciencias, Colombia;
MEYS CR, Czech Republic;
DNRF and DNSRC, Denmark;
IN2P3-CNRS and CEA-DRF/IRFU, France;
SRNSFG, Georgia;
BMBF, HGF and MPG, Germany;
GSRI, Greece;
RGC and Hong Kong SAR, China;
ISF and Benoziyo Center, Israel;
INFN, Italy;
MEXT and JSPS, Japan;
CNRST, Morocco;
NWO, Netherlands;
RCN, Norway;
MEiN, Poland;
FCT, Portugal;
MNE/IFA, Romania;
MESTD, Serbia;
MSSR, Slovakia;
ARRS and MIZ\v{S}, Slovenia;
DSI/NRF, South Africa;
MICINN, Spain;
SRC and Wallenberg Foundation, Sweden;
SERI, SNSF and Cantons of Bern and Geneva, Switzerland;
MOST, Taiwan;
TENMAK, T\"urkiye;
STFC, United Kingdom;
DOE and NSF, United States of America.
In addition, individual groups and members have received support from
BCKDF, CANARIE, Compute Canada and CRC, Canada;
PRIMUS 21/SCI/017 and UNCE SCI/013, Czech Republic;
COST, ERC, ERDF, Horizon 2020 and Marie Sk{\l}odowska-Curie Actions, European Union;
Investissements d'Avenir Labex, Investissements d'Avenir Idex and ANR, France;
DFG and AvH Foundation, Germany;
Herakleitos, Thales and Aristeia programmes co-financed by EU-ESF and the Greek NSRF, Greece;
BSF-NSF and MINERVA, Israel;
Norwegian Financial Mechanism 2014-2021, Norway;
NCN and NAWA, Poland;
La Caixa Banking Foundation, CERCA Programme Generalitat de Catalunya and PROMETEO and GenT Programmes Generalitat Valenciana, Spain;
G\"{o}ran Gustafssons Stiftelse, Sweden;
The Royal Society and Leverhulme Trust, United Kingdom.
 
The crucial computing support from all WLCG partners is acknowledged gratefully, in particular from CERN, the ATLAS Tier-1 facilities at TRIUMF (Canada), NDGF (Denmark, Norway, Sweden), CC-IN2P3 (France), KIT/GridKA (Germany), INFN-CNAF (Italy), NL-T1 (Netherlands), PIC (Spain), ASGC (Taiwan), RAL (UK) and BNL (USA), the Tier-2 facilities worldwide and large non-WLCG resource providers. Major contributors of computing resources are listed in Ref.~\cite{ATL-SOFT-PUB-2023-001}.

% End of text imported from the .//acknowledgements/Acknowledgements.tex input file

\clearpage

\printbibliography
 
\clearpage
% ATLAS Collaboration author list
% Reference date of HIGG-2020-18 is 2022-09-26
% Author list last updated on date 04-JUL-23
% Data extracted on 04-Jul-2023 for paper reference HIGG-2020-18
% at 2:11pm
 
\begin{flushleft}
\hypersetup{urlcolor=black}
{\Large The ATLAS Collaboration}

\bigskip

\AtlasOrcid[0000-0002-6665-4934]{G.~Aad}$^\textrm{\scriptsize 102}$,
\AtlasOrcid[0000-0002-5888-2734]{B.~Abbott}$^\textrm{\scriptsize 120}$,
\AtlasOrcid[0000-0002-1002-1652]{K.~Abeling}$^\textrm{\scriptsize 55}$,
\AtlasOrcid[0000-0002-8496-9294]{S.H.~Abidi}$^\textrm{\scriptsize 29}$,
\AtlasOrcid[0000-0002-9987-2292]{A.~Aboulhorma}$^\textrm{\scriptsize 35e}$,
\AtlasOrcid[0000-0001-5329-6640]{H.~Abramowicz}$^\textrm{\scriptsize 151}$,
\AtlasOrcid[0000-0002-1599-2896]{H.~Abreu}$^\textrm{\scriptsize 150}$,
\AtlasOrcid[0000-0003-0403-3697]{Y.~Abulaiti}$^\textrm{\scriptsize 117}$,
\AtlasOrcid[0000-0003-0762-7204]{A.C.~Abusleme~Hoffman}$^\textrm{\scriptsize 137a}$,
\AtlasOrcid[0000-0002-8588-9157]{B.S.~Acharya}$^\textrm{\scriptsize 69a,69b,p}$
% ...
\bigskip
\\

% ...
$^{29}$Physics Department, Brookhaven National Laboratory, Upton NY; United States of America.\\
% ...
$^{35}$$^{(a)}$Facult\'e des Sciences Ain Chock, R\'eseau Universitaire de Physique des Hautes Energies - Universit\'e Hassan II, Casablanca;$^{(b)}$Facult\'{e} des Sciences, Universit\'{e} Ibn-Tofail, K\'{e}nitra;$^{(c)}$Facult\'e des Sciences Semlalia, Universit\'e Cadi Ayyad, LPHEA-Marrakech;$^{(d)}$LPMR, Facult\'e des Sciences, Universit\'e Mohamed Premier, Oujda;$^{(e)}$Facult\'e des sciences, Universit\'e Mohammed V, Rabat;$^{(f)}$Institute of Applied Physics, Mohammed VI Polytechnic University, Ben Guerir; Morocco.\\
% ...
$^{55}$II. Physikalisches Institut, Georg-August-Universit\"{a}t G\"ottingen, G\"ottingen; Germany.\\
% ...
$^{69}$$^{(a)}$INFN Gruppo Collegato di Udine, Sezione di Trieste, Udine;$^{(b)}$ICTP, Trieste;$^{(c)}$Dipartimento Politecnico di Ingegneria e Architettura, Universit\`a di Udine, Udine; Italy.\\
% ...
$^{102}$CPPM, Aix-Marseille Universit\'e, CNRS/IN2P3, Marseille; France.\\
% ...
$^{117}$Department of Physics, New York University, New York NY; United States of America.\\
% ...
$^{120}$Homer L. Dodge Department of Physics and Astronomy, University of Oklahoma, Norman OK; United States of America.\\
% ...
$^{137}$$^{(a)}$Departamento de F\'isica, Pontificia Universidad Cat\'olica de Chile, Santiago;$^{(b)}$Millennium Institute for Subatomic physics at high energy frontier (SAPHIR), Santiago;$^{(c)}$Instituto de Investigaci\'on Multidisciplinario en Ciencia y Tecnolog\'ia, y Departamento de F\'isica, Universidad de La Serena;$^{(d)}$Universidad Andres Bello, Department of Physics, Santiago;$^{(e)}$Instituto de Alta Investigaci\'on, Universidad de Tarapac\'a, Arica;$^{(f)}$Departamento de F\'isica, Universidad T\'ecnica Federico Santa Mar\'ia, Valpara\'iso; Chile.\\
% ...
$^{150}$Department of Physics, Technion, Israel Institute of Technology, Haifa; Israel.\\
$^{151}$Raymond and Beverly Sackler School of Physics and Astronomy, Tel Aviv University, Tel Aviv; Israel.\\
% ...
$^{a}$ Also Affiliated with an institute covered by a cooperation agreement with CERN.\\
$^{b}$ Also at An-Najah National University, Nablus; Palestine.\\
$^{c}$ Also at Borough of Manhattan Community College, City University of New York, New York NY; United States of America.\\
$^{d}$ Also at Bruno Kessler Foundation, Trento; Italy.\\
$^{e}$ Also at Center for High Energy Physics, Peking University; China.\\
$^{f}$ Also at Center for Interdisciplinary Research and Innovation (CIRI-AUTH), Thessaloniki ; Greece.\\
% ...
$^{p}$ Also at Department of Physics, King's College London, London; United Kingdom.\\
% ...

\end{flushleft}

% Created with Glance <Atlas.Glance@cern.ch>

\end{document}